\documentclass[12pt]{article}

\usepackage{geometry}
\pdfoutput=1

\usepackage{amsmath,amssymb,amsfonts,amsthm,amscd,mathrsfs}
\usepackage{bm}
\usepackage{dsfont}
\usepackage{bbm}
\usepackage{esvect}
\usepackage{extarrows}
\usepackage{empheq}
\usepackage{cancel}

\numberwithin{equation}{section}

\usepackage{latexsym}
\usepackage{yfonts}

\usepackage{graphicx}
\usepackage{subcaption}
\usepackage{float}

\usepackage{tikz}
\usepackage{tikz-feynman}
\tikzfeynmanset{compat=1.1.0}

\usepackage{enumitem}
\setlist{itemsep=3pt, topsep=4pt}

\usepackage[margin=10pt,font=small,labelfont=bf]{caption}

\usepackage{xcolor}
\definecolor{darkblue}{rgb}{0.1,0.1,.7}
\definecolor{newyellow}{RGB}{128, 10, 7}

\usepackage[colorlinks=true, 
linkcolor=teal!75!black,
urlcolor=teal!75!black,
citecolor=newyellow,
]{hyperref}

\usepackage{cite}

\usepackage{xspace}
\usepackage{soul}
\usepackage{ifthen}
\usepackage[utf8]{inputenc}
\usepackage{setspace}

\definecolor{avocadogreen}{RGB}{86,130,3}

\newcommand{\tj}{\bar \jmath_0}

\newcommand{\cO}{\mathcal O}

\newcommand{\reef}[1]{(\ref{#1})}
\newcommand{\be}{\begin{equation}}
\newcommand{\ee}{\end{equation}}
\newcommand{\bea}{\begin{eqnarray}}
\newcommand{\eea}{\end{eqnarray}}
\newcommand{\ba}{\begin{equation}\begin{aligned}}
\newcommand{\ea}{\end{aligned}\end{equation}}

\newcommand{\Dg}{\Delta_{\mbox{\tiny gap}}}

\newcommand{\ud}{\mathrm d}
\newcommand{\mbf}{\mathbf}

\newcommand{\Df}{{\Delta_\phi}}

\def\D{\Delta}

\usepackage{bm}

\usepackage{cancel}

\def\XXint#1#2#3{{\setbox0=\hbox{$#1{#2#3}{\int}$}
     \vcenter{\hbox{$#2#3$}}\kern-.5\wd0}}

\newcommand*\pFqskip{8mu}
\catcode`,\active
\newcommand*\pFq{\begingroup
        \catcode`\,\active
        \def ,{\mskip\pFqskip\relax}%
        \dopFq
}

\catcode`\,12
\def\dopFq#1#2#3#4#5{%
        {}_{#1}F_{#2}\biggl[\genfrac..{0pt}{}{#3}{#4};#5\biggr]%
        \endgroup
}
\catcode`\,12

\begin{document}
\thispagestyle{empty}

\vspace*{-1.2cm}
\begin{center}
{\Large\bf
Super Sum rules for Long-Range Models
}
\end{center}

\vspace{0.8cm}

\begin{center}
{\bf
Kausik Ghosh$^{\tilde{\beta}_0}$,
Miguel F.~Paulos$^{\tilde{\omega}_1}$,
No\'e Suchel$^{\tilde{\omega}_2}$,
Zechuan Zheng$^{\tilde{\omega}_3}$
}\\[0.6cm]

{\small\em
${}^{\tilde{\beta}_0}$ Department of Mathematics, King’s College London\\
Strand, London WC2R 2LS, United Kingdom
}\\[0.3cm]

{\small\em
${}^{\tilde{\omega}_1,\tilde{\omega}_2}$ Laboratoire de Physique, Institut Philippe Meyer,\\
\'Ecole Normale Sup\'erieure, Universit\'e PSL, CNRS,\\
Sorbonne Universit\'e, Universit\'e Paris Cit\'e,\\
24 rue Lhomond, F-75005 Paris, France
}\\[0.3cm]

{\small\em
${}^{\tilde{\pi}}$ Perimeter Institute for Theoretical Physics,\\
Waterloo, ON N2L 2Y5, Canada
}
\end{center}

\vspace{0.4cm}

\begin{center}
{\small
\texttt{${}^{\tilde{\beta}_0}$kau.rock91@gmail.com}\quad
\texttt{${}^{\tilde{\omega}_1}$miguel.paulos@ens.fr}\quad
\texttt{${}^{\tilde{\omega}_2}$noe.suchel@ens.fr}\quad
\texttt{${}^{\tilde{\omega}_3}$zechuan.zheng.phy@gmail.com}
}
\end{center}

\vspace{0.8cm}

\begin{abstract}
We study sum rules that control the Regge limit of one-dimensional conformal
field theory (CFT) correlators and relate them to dual bulk scattering processes
at high energies in $\mathrm{AdS}_2$. By imposing the condition that no scattering
takes place in the bulk, these sum rules single out special solutions to crossing
symmetry that describe long-range models, which can be understood as free fields
in AdS with boundary interactions tuned to criticality. We test these sum rules
perturbatively in several distinct theories, namely the 1d long-range versions of the Ising,  $O(N)$ and Lee--Yang models, and find that they
correctly predict the CFT data characterising these theories.
Along the way we compute for the first time the leading contributions of quadruple-twist operators to the long range Ising correlator and
analyse their role in the new sum rules. Finally, we explore the consequences of imposing these
sum rules in a numerical bootstrap framework and find that they lead to
substantial reductions in the allowed parameter space.
\end{abstract}
\newpage
{
		\setlength{\parskip}{0.05in}
		\tableofcontents
		\renewcommand{\baselinestretch}{1.0}\normalsize
	}
	\setlength{\parskip}{0.08in}
    \setlength{\parindent}{0pt}
 \setlength{\abovedisplayskip}{14pt}
 \setlength{\belowdisplayskip}{14pt}
 \setlength{\belowdisplayshortskip}{10pt}
 \setlength{\abovedisplayshortskip}{10pt}
	\bigskip \bigskip

\section{Introduction}
	The conformal bootstrap uses basic principles of locality and unitarity to constrain large swathes of the space of possible CFT data \cite{Rattazzi:2008pe,Poland:2018epd}. Since so very little is input, the results have very broad applicability. However, one is often interested in constraining the parameters of a specific CFT. In this case, either the desired theory happens to saturate (or be close to) a bound/sits on a kink/lives on an island \cite{ElShowk:2012, Chang:2024whx} (in order of increasing luck) or one is forced to make more or less well motivated assumptions to get reasonable results.

    In this work we begin to explore how this state of affairs may be improved in the context of 1d CFTs. These theories describe observables of QFT in spaces which include an AdS$_2$ factor. Every mass scale $M$ in the QFT becomes a dimensionless modulus $M R_{\mbox{\tiny AdS}}$ in the space of 1d CFTs. Accordingly the problem described above is particularly acute here. Concretely, consider the theory of a single scalar in AdS$_2$. There are an infinite number of relevant interactions that may be written down:
    \ba
    \mathcal S=\int_{\mbox{\tiny AdS}}\,\bigg[ (\nabla \Phi)^2+m^2  \Phi^2+R^2 \sum_n g_n \Phi^n\bigg]
    \ea
    The question we are interested in is, how can we impose specific couplings $g_n$ from the perspective of the boundary CFT$_1$? 
    
    We propose that this can be done by considering scattering experiments in AdS$_2$. Since the interactions above are all relevant, then we expect that in the high energy limit they can be treated perturbatively, possibly after renormalisation. In particular we expect that the 2-to-2 S-matrix of the fundamental scalar admits an expansion of the schematic form:
    \ba
    T(s):=-i s [S(s)-1]\underset{s\gg 1}{\sim}  g_4^{\mbox{\tiny eff}}+o(1)
    \ea
    where $g_4^{\mbox{\tiny eff}}$ is an effective renormalised quartic coupling at high energies (at tree level $g_4^{\mbox{\tiny eff}}=g_4$). Other couplings are accessed in principle from subleading terms or higher-point scattering observables.
    The point now is that we have good control over how 2d S-matrices are encoded in the 1d CFT data \cite{QFTinAdS,Komatsu:2020sag,Ghosh:2025sic}. In particular such an S-matrix is reproduced by a CFT correlator containing a tower of `double trace' states with dimensions
    \ba
    \Delta_n=2\Delta_\phi+2n+\gamma_n,\,, \qquad  n^2 \gamma_n\underset{n\to \infty}\sim g_4^{\mbox{\tiny eff}}
    \ea
    Using crossing, we will show that the large $n$-behaviour of the $\gamma_n$ is controlled by a {\em super-bounded} sum rule  (or super sum rule for short), leading to
    \ba
    \sum_{\Delta} \lambda_{\phi\phi \Delta}^2 \tilde \beta_0(\Delta)\propto g_4^{\mbox{\tiny eff}}\label{sumrule}
    \ea
    where the precise proportionality factor can be worked out, and the $\tilde \beta_0$ functional has appeared before in the literature \cite{Mazac:2018ycv}.
    
       By itself this sum rule is perhaps only mildly interesting, as it can be thought of as computing the parameter $g_4^{\mbox{\tiny eff}}$ from the CFT data. 
       There is one important exception though, which is when we set those couplings to zero. This is a necessary condition for theories where the bulk lagrangian corresponds to a free field.  This does not necessarily mean a trivial theory: it is still possible to include boundary interactions which  can be tuned to lead to an interacting fixed point. Such theories are known as long-range models, and they correspond to mean free fields with local interactions. The most studied example is the long-range Ising model \cite{Fisher:1972zz,Paulos:2015jfa, Behan:2017dwr,Behan:2017emf,Behan:2018hfx,Giombi:2019enr,Behan:2023ile,Benedetti:2024wgx,Benedetti:2025nzp}, which in $d=1$ is described by an action of the form:
    \ba
    \mathcal S= \int \ud x\bigg[-\phi\, \Box^\alpha\, \phi + \lambda \phi^4\bigg],
    \ea
    where the kinetic term involves a laplacian raised to a fractional power (and is hence non-local). Such a field can be equivalently understood as the boundary value of a free field in AdS$_2$ with appropriately chosen mass, namely
    \ba
    m^2=\Df(\Df-1)\,, \qquad \Df=\frac{1-2\alpha}2
    \ea
    By tuning $\lambda$ we can reach a non-trivial CFT in the range $\frac 14<\alpha<\frac 12 \Leftrightarrow 0<\Df<\frac 14$. 

    In this paper we propose that long range models should satisfy the above sum rule with zero right-hand side. We will examine this claim in perturbation theory for three distinct long-range models: Ising, Lee-Yang and $O(N)$. These are long-range versions of the familiar local models, which by judicious choices of the parameter $\alpha$ can now lead to CFTs in any $d$ including $d=1$. For instance the (non-unitary)
    long range Lee-Yang model has a cubic interaction and leads to a CFT in the range $\Df<\frac 13$ and $\alpha<\frac 16$. 
    As for the $O(N)$ model, it has the same lagrangian as above but with the field $\phi$ now promoted to a vector of $O(N)$, $\phi\to \phi^i$ and $\phi^4\to (\phi^2)^2$. In this case we expect not one but actually two super sum rules, controlling two quartic AdS couplings :
    \ba
    \int_{\mbox{\tiny AdS}}\left[ g_4^{(1)} (\Phi^2)^2 +g_4^{(2)} (\Phi_i \overset{\leftrightarrow}{\nabla} \Phi_j)^2\right]
    \ea
     The checks are highly non-trivial: the sum rules can require regularisation as they do not always converge. After this is done, in all cases we find that the sum rules together with crossing are enough to fix all CFT data in perturbation theory to their correct values, i.e. those following from direct computation with Feynman diagrams. 
     
     Since the S-matrix is a function of $s$, one might expect there to be not just one or two but an infinite set of sum rules, which roughly correspond to its large$-s$ expansion. Indeed we will find evidence that this is the case, although such sum rules become more and more difficult to use due to harder and harder UV behaviour. A non-trivial computable example will arise for the $O(N)$ model, where a third functional arises on the same footing as the two previously mentioned, and whose sum-rule is also non-trivially satisfied.
     
     After these analytic checks, we will also explore the consequences of imposing the sum rule~\reef{sumrule} on the general space of CFT correlators, including it together with the usual crossing constraints and unitarity to derive new stronger bounds on CFT data. One outcome is that for $\Df\geq \frac 12$ the usual gap maximization bound goes from $\Delta_g\leq 1+2\Df$ to $\Delta_g\leq 2\Df$, so that it is now saturated by a generalized free boson rather than a generalized free fermion correlator \cite{Gaiotto:2013nva,Mazac:2016qev}. For $\Df<\frac 12$ instead the bound is $\Dg\sim 1$ and is saturated by a certain interacting solution. We also derive a universal bound on the leading irrelevant scalar. Unfortunately and in spite of the perturbative matching none of the numerical bounds we consider is actually saturated by the long range Ising model. 

    The outline of this paper is as follows. 
    
    \begin{itemize}
    \item In the next section we motivate the super sum rule. We discuss transparency of CFT correlators and how this constrains the OPE. In particular, we find under suitable conditions on the Regge limit that the sum rule above has to be satisfied with $g_4=0$. We then show that the same sum rule is also responsible for computing the constant piece in the scattering amplitude in the flat space limit of CFTs arising from QFTs in AdS.

    \item In section 3 we examine the consequences of the super sum rules and crossing symmetry for CFTs perturbatively close to GFFs. This includes the long range Ising, Lee-Yang and $O(N)$ models mentioned above. The super sum rules allow us to fix all perturbative data which is matched with exact diagrammatic computations. We also compute the leading contribution of quadruple-twist operators to the correlator of identical fundamental fields in the Ising model, which provides the necessary input for a future computation of the four-loop anomalous dimensions of double-trace operators.

    \item In section 4 we perform a numerical exploration of the consequences of the sum rule given above. We find large reductions in the allowed space but argue that further input is necessary to isolate the long range Ising model.
    
    \item Finally, we conclude with a discussion and outline directions for future work. Several technical appendices complement the main text.

\end{itemize}

   \section{Super sum rules}
   \subsection{The missing functional}
   We beginning by setting our conventions.  We write 1d CFT correlators as
    \ba
    \mathcal G(z):=\langle \phi(\infty)\phi(1) \phi(z)\phi(0)\rangle
    \ea
    and their OPE decomposition as
    \ba
    \mathcal G(z)=\sum_{\Delta} a_{\Delta} G_{\Delta}(z)\,, \quad G_{\Delta}(z)=z^{\Delta-2\Df}\, _2F_1(\Delta,\Delta,2\Delta,z)
    \ea
    with $a_{\Delta}:=\lambda^2_{\phi \phi \mathcal O_{\Delta}}$ related to the OPE coefficients. We recall that crossing symmetry is equivalent to the sum rules \cite{Mazac:2019shk} (see also \cite{Mazac:2016qev,Mazac:2018mdx})
    \ba
    \mathcal G(z)=\mathcal G(1-z) \quad \Longleftrightarrow \quad \left\{\begin{array}{ll}
    \sum_{\Delta} a_{\Delta} \beta_n(\Delta)=0\,, & n\geq 1\vspace{0.2cm} \\\sum_{\Delta} a_{\Delta} \alpha_n(\Delta)=0\,,& n\geq 0
    \end{array}
    \right. \label{eq:crossing}
    \ea
    for functionals satisfying the duality conditions
    \ba
    \beta_n(\Delta_m^{\tt gff})&=0\,,& \qquad \partial_{\Delta} \beta_n(\Delta_m^{\tt gff})&=\delta_{n,m}-e_n \delta_{m,0}\\
    \alpha_n(\Delta_m^{\tt gff})&=\delta_{n,m}\,,& \qquad \partial_{\Delta} \alpha_n(\Delta_m^{\tt gff})&=\delta_{n,m}-f_n \delta_{m,0}
    \ea
    with $m\geq 0$, $\Delta_m^{\tt gff}:=2\Df+2m$, and $e_n,f_n$ known coefficients. Note the conspicuous absence of the functional $\beta_0$ in the list of sum rules above (in particular leading to the awkward duality relations for $m=0$).

 We are interested in CFTs which arise from QFTs of a scalar field $\Phi$ in AdS$_2$ deformed by relevant interactions. In such theories we can produce 1d CFT correlators from bulk ones by the push map, $\Phi(y,x)\sim y^{\Df} \phi(x)$, with small $y$ corresponding to the AdS boundary. In the pure free bulk limit the CFT is a generalized free field, and we have
    \ba
    \mathcal G_{\tt gff}(z)=1+\frac 1{z^{2\Df}}+\frac 1{(1-z)^{2\Df}}=\frac 1{z^{2\Df}}+\sum_{n=0}^\infty a_{n}^{\tt gff}\, G_{\Delta_n}(z) \label{gffcorr}
    \ea
    with
    \ba
    a_n^{\tt gff}=a_{\Delta_n^{\tt gff}}^{\tt gff}\,, \qquad a_{\Delta}^{\tt gff}=\frac{\sqrt{\pi } 2^{3-2 \Delta } \Gamma (\Delta ) \Gamma \left(\Delta +2 \Delta _{\phi }-1\right)}{\Gamma \left(\Delta -\frac{1}{2}\right) \Gamma \left(\Delta -2 \Delta _{\phi }+1\right) \Gamma
   \left(2 \Delta _{\phi }\right)^2}\:.
    \ea
    
    The absence of  the $\beta_0$ functional is intimately related to the existence of the $\Phi^4$ term in AdS$_2$~\cite{Mazac:2018ycv}. Indeed, consider deforming the GFF solution without introducing additional states, allowing for small changes of the OPE coefficients anomalous dimensions $\gamma_n=\Delta_n-\Delta_n^{\tt gff}$. Crossing as formulated in \reef{eq:crossing} uniquely fixes the deformation, giving
    \ba
    \gamma_n=2\gamma_0 \frac{e_n}{a_n^{\tt gff}}\underset{n\gg 1}\approx \frac{\gamma_0}{n^2}
    \ea    
    Such behaviour corresponds to a contact interaction in the flat space limit, and in fact it is possible to compute the $\gamma_n$ exactly and check that they precisely match with the result of switching on a $\Phi^4$ deformation in the bulk. Such a deformation could have been ruled out had there been an additional sum rule for a functional which we denote $\tilde \beta_0$ satisfying
    \ba
    \tilde \beta_0(\Delta_m^{\tt gff})=0\,, \qquad \partial_{\Delta}\tilde \beta_0(\Delta_m^{\tt gff})=\delta_{m,0}
    \ea
    Such a functional can in fact be constructed \cite{Mazac:2018ycv}, but it does not satisfy the `swapping' property~\cite{Qiao:2017lkv} for ordinary CFT correlators containing the identity. In very practical terms this means that there exist solutions to crossing where the $\tilde \beta_0$ sum rule is not even convergent, let alone equal to zero. We give a precise formula for $\tilde \beta_0$ and discuss applicability of its corresponding sum rule in appendix \ref{app:beta0}. 
    
For CFT correlators close to a GFF in the UV (i.e. large $\Delta$) the spectrum localises near the double zeros of $\tilde \beta_0$. This leads to improved convergence of the sum rule and possibly a finite result. What should this result be? Well, somewhat trivially, for the simple perturbative example we just considered the sum over states just gives $\gamma_0$ which is proportional to the bulk coupling $g_4$. This can be promoted to the non-perturbative statement:    %

\setlength\fboxrule{1.2pt}
\setlength{\fboxsep}{8pt} 
{\hspace{-6pt}\noindent\fbox{\parbox{\textwidth-18pt}{
{\bf Super sum rule:}
    \ba
    \sum_\Delta a_{\Delta} \tilde \beta_0(\Delta)= \frac{g_4^{\mbox{\tiny eff}}}{16\pi \kappa_{\Df}}
    \ea
\vspace{-0.3cm}
}}}
   \vspace{0.3cm} 

\noindent
The precise proportionality constant depends on our precise bulk lagrangian conventions and is worked out in the appendix \ref{app:constant}. Note that the right-hand side can be independently defined via the large $z$ behaviour of the CFT correlator as explained in appendix~\ref{app:beta0}, roughly $\mathcal G(z)-1\sim g_4^{\mbox{\tiny eff}}/z$.
    
    Our proposal is then that the $\tilde \beta_0$ sum rule, at least when convergent, measures the effective $\Phi^4$ coupling in AdS$_2$, even when this coupling is large, the idea being that since its a relevant deformation, in the UV it can always be thought of as weak.  As the coupling increases, low energy states pick up $O(1)$ anomalous dimensions and the sum on the left takes longer to converge, consistent with the increased right hand side.

    On its own the sum rule above is of limited use, as we can think of it as computing the a priori unknown right hand side. But there is one exception, namely if we are interested in theories with a free bulk description. In this case the righthand side is simply zero and we end up with a new non-trivial constraint on the CFT data. 
    
    \subsection{Transparency and the super sum rule}
    \label{sec:transparency}

    Let us now discuss more carefully what we mean by CFT correlators which are asymptotically free, or GFF like, and how this relates to the $\tilde \beta_0$ sum rule. Our main point here is that
  the sum rule can make sense even for a correlator which is not super Regge-bounded, in particular for $\mathcal G(z)\sim 1$, as long as its double discontinuity is.    
    We should note that the discussion here overlaps and refines the one in \cite{Paulos:2020zxx}. There are also notable similarities with the higher-D analysis in \cite{Caron-Huot:2020ouj}.

    We begin by splitting CFT correlators into two classes:
    \ba
    \mbox{Transparent}\Longrightarrow \qquad & \lim_{z \to \infty} |\mathcal G(z)|&=1&\\
    \mbox{Opaque} \Longrightarrow \qquad & \lim_{z \to \infty} |\mathcal G(z)|&<1&
    \ea
    The free GFF correlator \reef{gffcorr} is transparent and intuitively we expect that transparency should not be affected by adding relevant interactions, at least for sufficiently small (but finite) coupling. Accordingly we may hope that any such correlator should have an OPE density which becomes GFF-like in the UV.
   
    In appendix \ref{app:freedom} we show that this is indeed the case under sufficiently strong assumptions on the Regge limit of the correlator. Concretely define the double discontinuity:
     \ba
    \ud^2 \mathcal G(z):=(1-z)^{2\Df}\mathcal G(z)-\mathcal R_z\mathcal G(\tfrac{z}{z-1})\,.
    \ea
    Transparency implies that the double discontinuity decays as $z\to 1$.
    Then we prove that
    \ba
    \ud^2 \mathcal G(z)\underset{z\to 1}\sim \nu (1-z)^{1-\tj} \quad \Longrightarrow\quad \langle \gamma_n^2\rangle=O(n^{2\tj-1})
    \ea
where we have set 
\ba
\langle \gamma_n^2\rangle:= \sum_{|\Delta-\Delta_n^{\tt gff}|\leq 1} \frac{4}{\pi^2} \left(\frac{a_{\Delta}}{a_{\Delta}^{\tt gff}}\right)\sin[\tfrac{\pi}2(\Delta-\Delta_n)^2]\,.
\ea
 For $\tj<\frac 12$ the above implies that anomalous dimensions concentrate at the GFF values assuming 
\ba
c'<\lim_{n\to \infty} \sum_{|\Delta-\Delta_n^{\tt gff}|\leq 1} \left(\frac{a_{\Delta}}{a_{\Delta}^{\tt gff}}\right)<c\,.
\ea
In turn this has been shown to hold in a variety of cases in \cite{Mazac:2018ycv} (and conjectured that $c=c'=1$).

Now that we know that the spectrum is asymptotically free, we can use crossing under the form of the sum rules \reef{eq:crossing} to get more fine grained information. Concretely, consider the $\beta_n$ sum rules. We begin by the trivial rearrangement
\ba
\sum_\Delta a_{\Delta} \beta_n(\Delta)=0 \Longleftrightarrow \sum_{\Delta> \Delta^*} a_{\Delta} \beta_n(\Delta)=-\sum_{0\leq \Delta\leq \Delta^*} a_\Delta \,\beta_n(\Delta)
\ea
Let us now take large $n\gg \Delta^*\gg 1$. On the left we can use the asymptotic form of the functionals
\ba
\beta_n(\Delta) \sim \frac{4}{\pi^2}\,\left(\frac{a_{n}^{\tt gff}}{a_{\Delta}^{\tt gff}}\right)\,\sin^2[\tfrac{\pi}2(\Delta-2\Df)] \,\frac{4 \Delta \Delta_n^2}{\Delta^4-\Delta_n^4} \label{asympbeta}
\ea
to write the sum as proportional to
\ba
\langle \gamma_n\rangle+O(n^{2\tj})-n^{-2}\, T_{\tj}(\Delta^*)\,, \quad 
\ea
This result is obtained by splitting the sum over states into a region parametrically close to~$\Delta_n$ and the remainder bulk sum. In the former we get
\ba
\langle \gamma_n \rangle=\sum_{|\Delta-\Delta_n|<\epsilon}\left(\frac{a_{\Delta}}{a_{\Delta}^{\tt gff}}\right) (\Delta-\Delta_n)\,, 
\ea
for some small $\epsilon>0$. Note that $\langle\gamma_n\rangle=O(n^{\tj-\frac 12})$. As for the remainder it is easy to see that the bulk sum is bounded as $O(n^{2\tj})$, but there can also be contributions arising mainly from the region near $\Delta=\Delta^*$ (which can become dominant if $\tj$ is too negative). The leading piece is a $n^{-2}$ term. Under suitable assumptions we can swap the large $n$ limit with the sum  over states to determine
\ba
T_{\tj}(x)=\sum_{\Delta>\Delta^*}\, \frac{4}{\pi^2}\left(\frac{a_{\Delta}}{a_{\Delta}^{\tt gff}}\right) \,\sin^2[\tfrac{\pi}2(\Delta-2\Df)] \,\Delta
\ea
This works as long as the above sum is finite, a sufficient condition for this being $\tj<-\frac 12$.

Turning to the right-hand side, using again asymptotic expansion for the functional action (fixed $\Delta$ , large $n$) we find it is given by
\ba
\frac{\kappa_{\Df} \sum_{\Delta\leq \Delta^*} a_\Delta \, \tilde \beta_0(\Delta)}{n^2}-a_{\Delta_g}\,\frac{r(\Delta_g)}{n^{2\Delta_g}}+\ldots \label{gamman}
\ea
where $\Delta_g$ is the dimension of the lowest non-identity operator and
\ba
\kappa_{\Df}=\frac 12\frac{\Gamma(2\Df)^4}{\Gamma(\Df)^4 \Gamma(4\Df-1)}\,, \qquad r(\Delta):=2^{-2\Delta} \left(\frac{\Gamma(2\Df)}{\Gamma(2\Df-\Delta)}\right)^2\, \frac{\tan(\tfrac{\pi}2 \Delta)}{\pi}\,
\ea
The sum over states in $\tilde \beta_0$ sum rule converges as long as $\tj<-\frac 12 $, in which case we can take the limit $\Delta^*\to \infty$ of the above. In fact to leading order in $\Delta^*$ we have
\ba
 \sum_{\Delta>\Delta^*} a_{\Delta} \tilde \beta_0(\Delta)\sim T_{\tj}(\Delta^*)/\kappa_{\Df} \label{eq:largebeta}
\ea
Thus we find
\ba
\langle \gamma_n\rangle+O(n^{2\tj})=\frac{\kappa_{\Df}}{n^2}\sum_{\Delta} a_{\Delta}\tilde \beta_0(\Delta)-a_{\Delta_g}\,\frac{r(\Delta_g)}{n^{2\Delta_g}} \label{assympgamman}
\ea
This is the main equation of this subsection. We will often use it ignoring the possible $O(n^{2\tj})$ piece on the left. Suppose $\tj<-1$. Then a simple computation shows that
\ba
\mathcal G(z)-1\underset{z\to \infty}= \frac{i  \pi}{(\Df-\tfrac 12)^2} \,\frac{\kappa_{\Df}\sum_{\Delta} a_{\Delta} \tilde \beta_0(\Delta)}z+\ldots
\ea
That is, the $1/z$ fall off in the correlator is controlled by the $\tilde \beta_0$ sum rule. This is consistent with our previous analysis of this functional controlling the $\Phi^4$ contact term in AdS, as such terms give indeed a $1/z$ fall-off to the correlator. Note that for $\tj<-\frac 32$
the bound $\langle \gamma_n\rangle=O(n^{\tj-\frac 12})$ implies that this term must be absent.

Let us recap what we have learned. Firstly, we have seen that appropriate bounds on the double discontinuity (i.e. transparency) imply that the CFT spectrum must become asymptotically GFF like. Using crossing this means that there can be a $1/z$ term in the correlator controlled by the $\tilde \beta_0$ sum over states. This sum is convergent under appropriate assumptions on the Regge behaviour $\tj$, and is even constrained to be zero if $\tj$ is sufficiently negative. 
Thus we learn that CFT correlators which have sufficiently Regge bounded double discontinuity have to satisfy additional sum rules \cite{Paulos:2020zxx}. We say sum rules plural, as looking at further subleading terms one finds other functionals controlling $1/z^3$, $1/z^5$, etc terms in the large $z$ expansion of the correlator. 

One important thing to note in this derivation is that the sum rule $\tilde \beta_0$ is not guaranteed to be convergent. However, we can still expect that for any $\tj$ the $1/z$ piece of the correlator is controlled by some regulated version of the $\tilde \beta_0$ sum rule. Note that the crucial step arises when extracting the $1/n^2$ term from the term $\sum_{\Delta>\Delta^*} a_{\Delta} \beta_n(\Delta)$ to obtain the tail piece $T_{\tj}(\Delta^*)$. In situations where we have some degree of analytic control over the spectrum then this tail piece can still be defined by analytic continuation. We will see this in practice below. 

\subsection{Bulk scattering and its absence}
     
We now further develop on the physical picture of the $\tilde \beta_0$ sum rule as measuring certain interactions in AdS$_2$. Consider a family of bulk 2d QFTs labelled by a dimensionless coupling $\Dg:=m_g R_{\mbox{\tiny AdS}}$ where $m_g$ is the typical mass scale of the QFT. Then the bulk S-matrix is given by the phase shift formula \cite{QFTinAdS}
     \ba
     S(s):=e^{i \gamma(s)}=\lim_{\Dg\to \infty} \frac{1}{(\Dg)^\alpha}\sum_{|\Delta-\Dg\sqrt{s}|<(\Dg)^{\alpha}} \left( \frac{a_{\Delta}}{a_{\Delta}^{\tt gff} }\right)\,e^{-i \pi (\Delta-2\Df)}
    \ea
    for some $0<\alpha<1$.
    For sufficiently large $s$ we have
    \ba
    T(s):=-i s (S(s)-1) \underset{s\to \infty} \sim s\,\gamma(s)=
    -\frac{4 \pi  n^2}{\Dg^2} \langle \gamma_n\rangle\bigg|_{n\sim \frac{\Dg\sqrt{s}}2}=-\frac{4\pi\kappa_{\Df}\sum_{\Delta} a_{\Delta} \tilde \beta_0(\Delta)}{\Dg^2}+\ldots
    \ea
     For a generic interacting QFT this is some $O(1)$ number and thus
    \ba
    \sum_{\Delta} a_{\Delta} \tilde \beta_0(\Delta)\propto \Dg^2 \propto g_4^{\mbox{\tiny eff}}
    \ea
    where $g_4^{\mbox{\tiny eff}}$ is an effective quartic interaction in the bulk, large in AdS units in the flat space limit. Indeed, such interactions lead to $\gamma_n\sim g_4^{\mbox{\tiny eff}}/n^2 \Rightarrow T(s)\sim O(1)$. The reason why above it is possible to get a possibly large number on the right hand side is because unlike ordinary functional sum rules, the $\tilde \beta_0$ sum rule only converges when anomalous dimensions decay, which can happen at parametrically large values of $\Delta$. 
    More generally, from the asymptotic expansion of $\langle \gamma_n\rangle$ we can extract higher subleading powers in $1/s$ of the S-matrix. By expanding the $\beta_n$ sum rule at large $n$ one can find
    \ba
    \langle \gamma_n\rangle= \sum_{k=0}^\infty c_k(n) \frac{ \Sigma_k}{n^{2+4k}}+\ldots \label{eq:largeh1}
    \ea
    where we have defined the supersums:
    \ba
    \Sigma_k:= \sum_{\Delta} a_{\Delta} \xi_k(\Delta) \label{eq:largeh2}\,.
    \ea
    The $\xi_k(\Delta)$ are chosen to satisfy
    \ba
    \xi_k(\Delta_{-p})=\delta_{p,k}\,, 1\leq p\leq k
    \ea
    after which choice the $c_k(n)$ are determined uniquely, and in particular $c_k(n)\underset{n\to \infty}=O(1)$. Note that%
    \ba
    \xi_0\propto \tilde \beta_0
    \ea
    The higher $k$ functionals $\xi_k$ have worse and worse convergence properties at large $\Delta$,
    \ba
    a_{\Delta}^{\tt gff} \xi_k(\Delta)\underset{\Delta \gg 1}\approx \frac{4}{\pi^2}\,\sin^2[\tfrac{\pi}2(\Delta-2\Df)]\times O(\Delta^{1+4k})
    \ea
    In the flat space limit this allows for the sums to be parametrically large $\Sigma_k=O(\Dg^{2+4k})$ in order to reproduce the large $s$ expansion of the S-matrix. After resumming all such terms, the $\beta_n$ equations become a dispersion relation for the $S$-matrix \cite{Ghosh:2025sic}.

    The conclusion is that $\tilde \beta_0$ as well as the higher functionals $\xi_k$ encode the high energy behaviour of the bulk S-matrix, at least when the CFT correlator is associated to a QFT in AdS. This suggests the following proposal:\vspace{0.5cm}\\
    {\bf Super sum rules}
    \begin{itemize}
    \item CFT correlators arising from a free bulk description satisfy the no-scattering super sum rules:
    \ba
    \Sigma_k:=\sum_{\Delta} a_{\Delta} \xi_k(\Delta)=0\,, \qquad k=0,\ldots,\infty
    \ea
    \end{itemize}
    In practice it is difficult to impose all such sum rules because of their bad convergence properties, which may require further and further UV subtractions.
    In this work we will investigate only the consequences of the first such sum rule.
    
\section{Perturbative checks}
    \subsection{Long range Ising}
    \label{sec:lri}
    The long range Ising CFT can be defined as the critical point of the theory
    \ba
    S=\int_{\mbox{\tiny AdS}} \left[ (\nabla \Phi)^2+m^2 \Phi^2\right]+\lambda \int_{\partial \mbox{\tiny AdS}}\, \phi^4
    \ea
    Here $\phi$ is the boundary limit of the bulk field $\Phi$. Note that its dimension is protected and determined as
    \ba
    m^2 R^2=\Df(\Df-1)
    \ea
    For $\Df<\frac 14$ the operator $\phi^4$ is relevant and the theory flows to a non-trivial fixed point. Setting $\epsilon=(1-4\Df)\ll 1$ it is possible to compute the data of this CFT perturbatively in $\epsilon$, {\em \`a la} Wilson-Fisher. Our aim is to reproduce the data of this perturbative expansion, or more precisely that of the correlator $\mathcal G=\langle \phi \phi \phi \phi\rangle$, using crossing symmetry together with the $\tilde \beta_0$ sum rule.

\begin{figure}[t]
  \centering
  \subfloat{\includegraphics[width=0.49\textwidth]{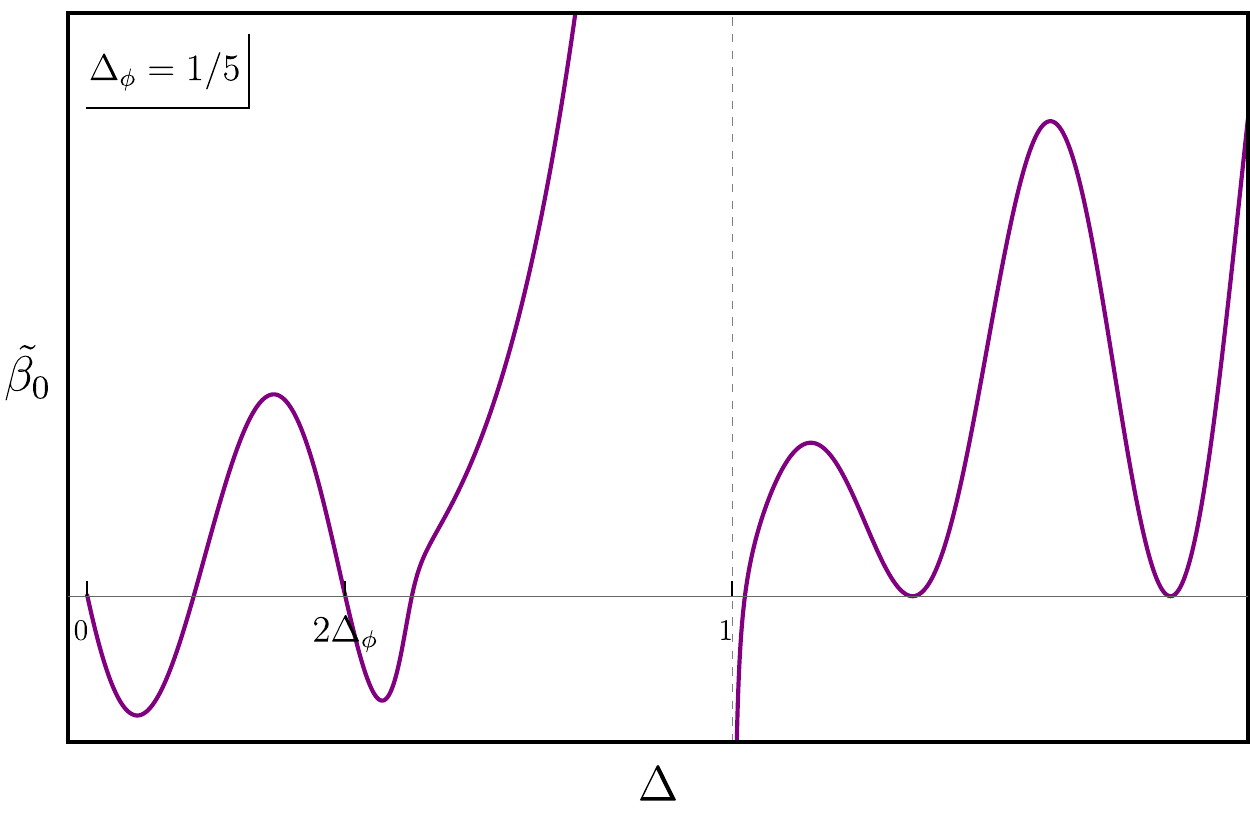}\hspace{0.3
  cm}\hfill \includegraphics[width=0.49\textwidth]{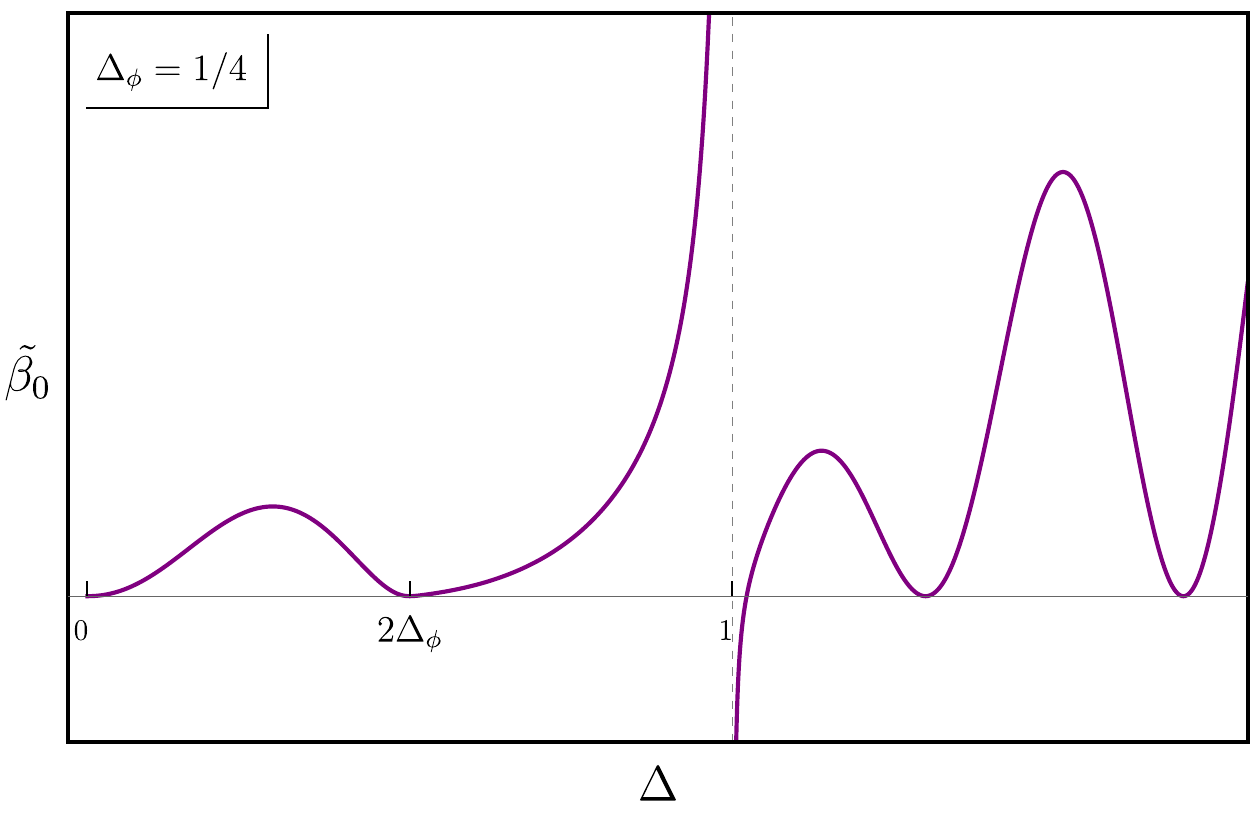}} \\
  \subfloat{\includegraphics[width=0.49\textwidth]{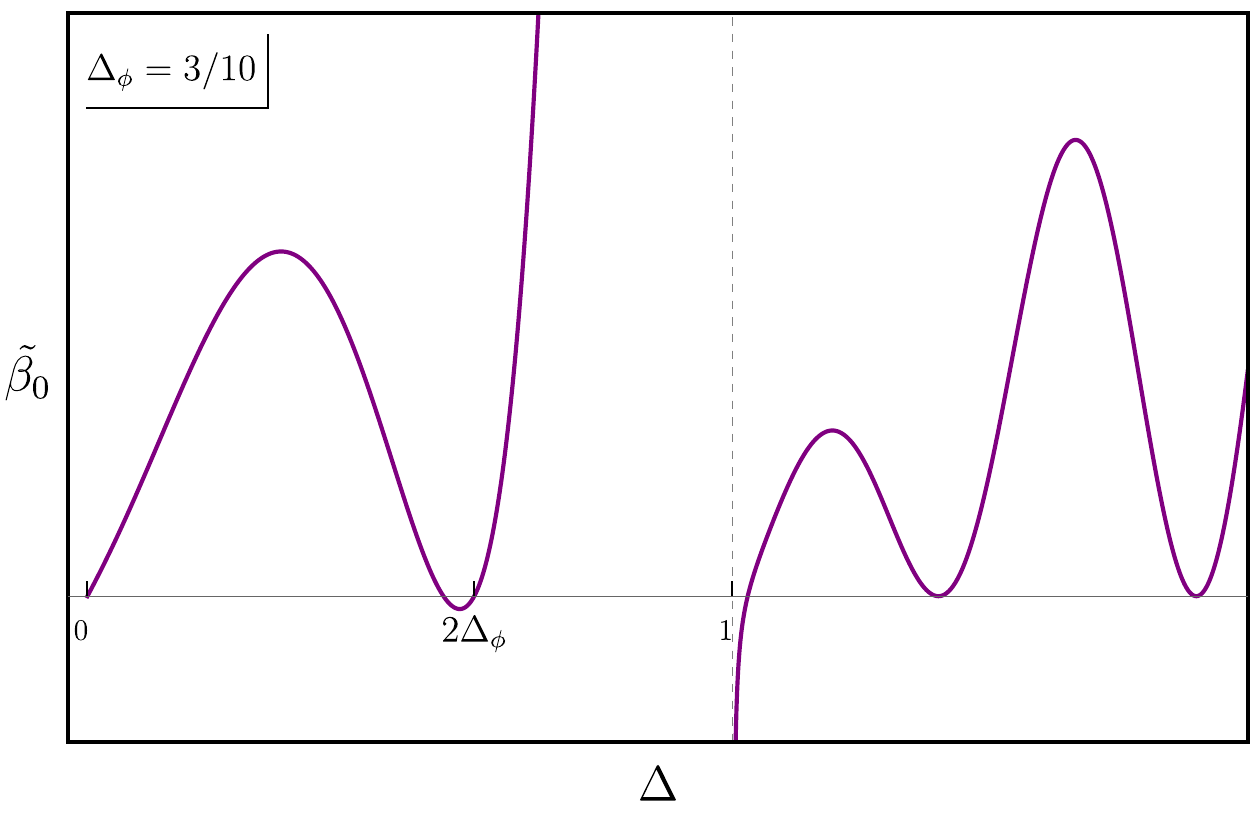}\hspace{0.3
  cm}\hfill \includegraphics[width=0.49\textwidth]{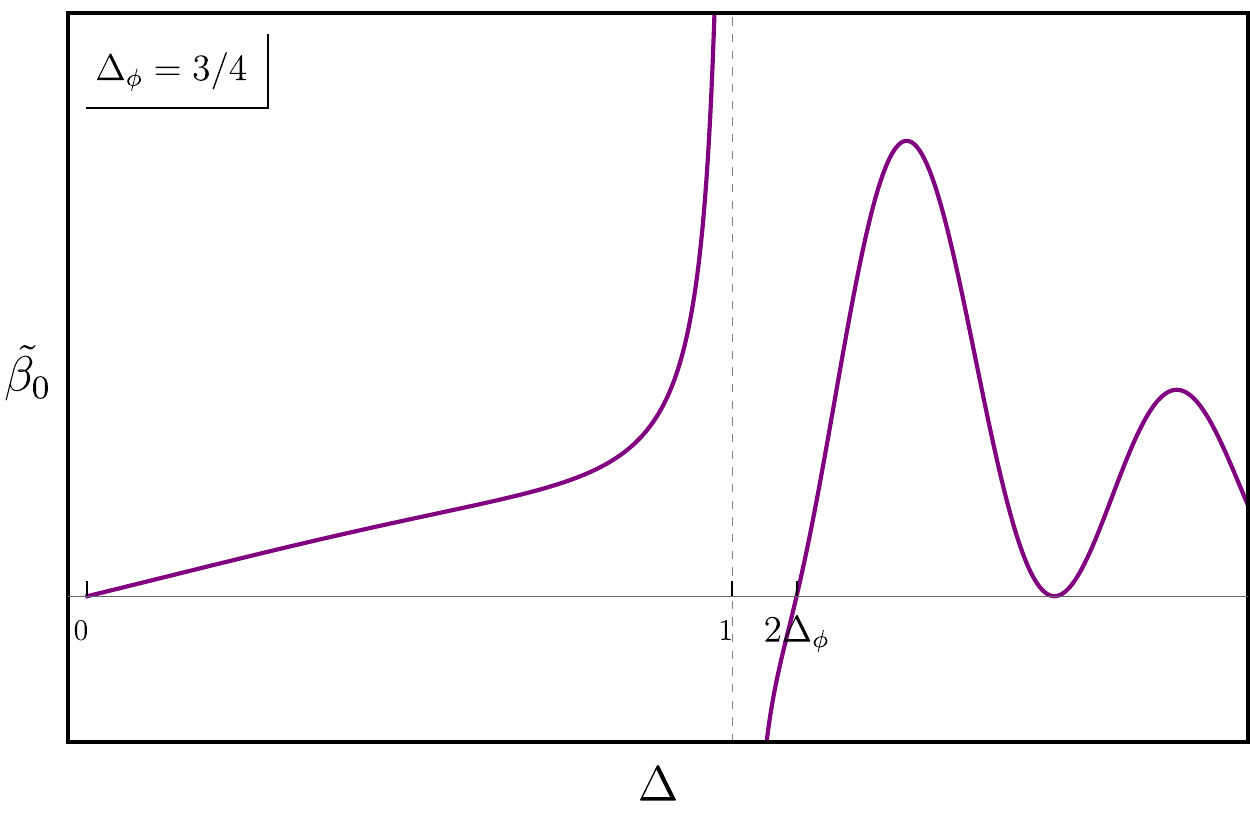}}
  \caption{
The functional $\tilde{\beta}_0(\Delta)$ is shown for $\Delta_{\phi}=1/5$ (top left), $\Delta_{\phi}=1/4$ (top right), $\Delta_{\phi}=3/10$ (bottom left) and $\Delta_{\phi}=3/4$ (bottom right). For clarity, the functionals have been rescaled.
Note the presence of a negative region just above $\Delta=1$.
}
  \label{fig: superbeta}
\end{figure}

    A first observation is that examination of the functional action shows that it blows up precisely at~$\Df=\frac 14$.  It is therefore convenient to work with the functional in a different normalisation,
    \ba
    \tilde \beta_0\to \hat \beta_0:=-\frac{\epsilon}4 \tilde \beta_0\,.
    \ea
    A plot of the renormalized functional action for different values of $\Df$ is shown in figure~\ref{fig: superbeta}.
    We see that $\Df=\frac 14$ is indeed a special value as the action develops a new zero in the vicinity of~$\Delta=2\Df$ (with which it coincides at exactly $\Df=\frac 14$, thus the blow up from having demanded $\partial_{\Delta} \tilde \beta_0(2\Df)=1$). This new zero signals the appearance of a new branch of solutions to crossing as we will now see.

    Consider then doing perturbation theory around the GFF solution \reef{gffcorr}. For a generic value of $\Df$ the crossing constraints have no solution. This is expected: the only such deformation would correspond as we've seen to the $\Phi^4$ coupling in AdS$_2$, which the super sum rule sets to zero. Remarkably however, this fails to be the case exactly in the vicinity of $\Df=\frac 14$.  To leading order in $\epsilon$ the $\beta_n$ equations set all anomalous dimensions to zero, $\gamma_n^{(1)}=0$. As for the $\hat \beta_0$ sum rule it gives
    \ba
    \hat \beta_0(0)+a_{\phi^2}\hat \beta_0(\Delta_{\phi^2})+\sum_{m=1}^\infty a_m \hat \beta_0(\Delta_m)
    \ea
    The functional $\hat \beta_0$ has double zeros for $\Delta_m=2\Df+2n$ with $n>1$ for any $\Df$. Let us set 
    \ba
    \Delta_{\phi^2}=2\Df+\epsilon \gamma_{0}^{(1)}+\epsilon^2 \gamma_{0}^{(2)}
    \ea
    and expand the above for small $\epsilon$:
	\bea
	-\gamma_0^{(1)}\, \partial_{\Df} \partial_{\Delta}\hat \beta_0(2\Df)\bigg|_{\Df=\frac 14}+2 \left(\gamma_0^{(1)}\right)^2 \partial_{\Delta}^2\hat \beta_0(2\Df)\bigg|_{\Df=\frac 14}=0
	\eea
    Direct evaluation of the functional action leads to
    \ba
    -\gamma_0^{(1)}+3 \left(\gamma_0^{(1)}\right)^2=0 \quad \Longrightarrow\quad  \gamma_0^{(1)}=0 \quad \mbox{or}\quad \gamma_0^{(1)}=\frac 1 3\,.
    \ea
    Let us now go to the next order. The $\beta_n$ equations determine anomalous dimensions which are $O(\epsilon^2)$. A detailed computation gives
    \ba
    \gamma_n^{(2)}&=-(a_n^{\tt gff})^{-1}\left(\gamma_{0}^{(1)}\right)^2\partial^2_{\Delta} \beta_n(2\Df)\bigg|_{\Df=\frac 14}+\frac 12 \left(\,\gamma^{(1)}_0 \partial_{\Df}[(a_n^{\tt gff})^{-1}\partial_{\Delta} \beta_n(2\Df)]\right)\bigg|_{\Df=\frac 14}\\
    &=-\frac{1}{9 n}\,.
    \ea
    Note that this is precisely consistent with \reef{assympgamman}.
    Now let us turn to the $\hat \beta_0$ equation. Let us assume that at this order only $\phi^2$ contributes. We will discuss this assumption shortly. We get:
    \ba
    \left[ \gamma_0^{(2)}-6\gamma_0^{(2)}\gamma_0^{(1)}-\frac 23 \left(\gamma_0^{(1)}\right) ^3\partial_{\Delta}^3\hat \beta_0(2\Df)+\frac 12 \left(\gamma_0^{(1)}\right) ^2 \partial_{\Df} \partial^2_{\Delta}\hat \beta_0(2\Df)\right]\bigg|_{\Df=\frac 14}=0
    \ea
    By numeric evaluation of $\hat \beta$ to high accuracy one finds
    \ba
    \gamma_0^{(2)}=\frac{1}{9}[\pi+\log(16)]:=\gamma_0^{LRI,(2)}
    \ea
    Overall we can write
    \ba
    \Delta_{\phi^2}=\frac 12-\frac{\epsilon}6+\frac{\epsilon^2}{9}[\psi(\tfrac 12)-2\psi(\tfrac 14)+\psi(1)]+O(\epsilon^3)
    \ea
    which precisely agrees with the perturbative data of the critical long-range Ising model in 1d \cite{Fisher:1972zz}.
    
    Now let us discuss further our assumption that no other states contribute to the sum rule. As it turns out this is not quite correct. Consider first the double trace operators. As their anomalous dimensions are $O(\epsilon^2)$ then individually they contribute to $\hat \beta_0$ as $O(\epsilon^4)$. However, the sum over all such operators turns out to be divergent. This leads to an enhancement which brings their contribution to $O(\epsilon^3)$. To compute this contribution we write
    \ba
    \gamma_n^{(2)}=-\frac{1}{9 n^{2\Delta_{\phi^2}}}
    \ea
    as expected from \reef{assympgamman}. Focusing on the divergent part of the sum over states (i.e. large $n$) we can use \reef{eq:largebeta} to find
    \ba
    \sum_n a_n^{\tt gff}\hat \beta_0(\Delta_n) \underset{n\gg 1}\sim \frac{\epsilon^4}{(-4\kappa_{\Df}/\epsilon)} \sum_{n} 2 n \left(\gamma_n^{(2)}\right)^2\sim -\frac{\Gamma(\tfrac 14)^ 4}{54 \pi^2}\,\epsilon^3
    \ea
    where we used zeta function regularization.
    Next, we must take into account the $\phi^4$ operator. We have
    \ba
    \Delta_{\phi^4}=1+\epsilon\,, \qquad \lambda_{\phi \phi \phi^ 4}^2=\frac{8 \pi^ 2}{27} \epsilon^4
    \ea
    Since $\hat \beta_0$ has a pole at $\Delta=1$ this again leads to an enhanced contribution,
    \ba
    \lambda_{\phi \phi \phi^4}^2 \hat \beta_0(\Delta_{\phi^4})\sim -\frac{\Gamma(\tfrac 14)^ 4}{54 \pi^2}\,\epsilon^3
    \ea
    Remarkably these contributions come out the same, but unfortunately with the same sign, so they do not cancel. 
    But this is not the end of the story, as we must consider the contributions from more general operators of the form $(\phi^4)_n$. Their individual OPE coefficients are again $O(\epsilon^4)$ (as for $n=0$), but their sum can in principle lead to a divergence in the sum rule which has to be regulated. In order to check this we need to compute these OPE coefficients. We do so in appendix \ref{Appendix: twist4} relying heavily on the OPE relations satisfied by the long range Ising model \cite{Paulos:2015jfa}. One finds
    \ba
    \frac{\lambda^2_{\phi \phi [\phi^4]_{n}}}{a^{\tt gff}_{1+2n}}\underset{n\gg 1}=\epsilon^4\left[\frac{a_1}{n^2}\left(n-\frac 14\right)+\frac{a_2}{n^2}+\ldots\right]
    \ea
    with
    \ba
    a_1=\frac{\pi^3}{162}\,, \quad a_2=\frac 23 \frac{ \pi^2}{54}
    \ea
    It can be checked that the powers are consistent with what is expected from the crossing equation for the $\langle \phi \phi \phi^2\phi^2\rangle$ in the s-channel in order to reproduce $\phi,\phi^3$ in the $t$-channel.
    
   These contributions can lead to a divergence in the supersum rule. Assume that to leading order the power law terms $1/n^k$ get corrected to $1/n^{k+\epsilon \nu_{k}}$ for some $\nu_k$. Focusing on the terms leading to a pole in $\epsilon$ and hence an enhancement, we find:
    \ba
    \sum_n \lambda^{2}_{\phi \phi [\phi^4]_{n}} \hat \beta_0(\Delta_{[\phi^4]_n})\underset{n\gg 1}\sim \frac{\epsilon^4}{(-4\kappa_{\Df}/\epsilon)}\, \sum_n \frac{4}{\pi^2} \frac{a_2}{n^{1+\nu_2 \epsilon}} \sim \frac{2/3}{\nu_2}\,2\, \frac{\Gamma(\tfrac 14)^4}{54\pi^2}\,\epsilon^3
    \ea
    where we used that $a^{\tt gff}_{1+2n} \hat \beta_0(1+2n)\propto (n+1/4+\ldots)$ so that the $a_1$ term actually does not contribute to this order.
    Thus overall cancellation is possible if $\nu_2=2/3$.
    The simplicity of this result after a highly non-trivial computation is for us encouraging of the validity of intermediate steps. It would be important in the future to determine the actual exponent in order to test our proposal.

    We have also done the following check. Rather than using the $\tilde \beta_0$ sum rule, we simply computed the double trace anomalous dimension to $O(\epsilon^3)$, i.e. $\gamma_n^{(3)}$ using the normal functionals, therefore as a function of $\gamma_0$. Expanding this quantity at large $n$ it is possible to express it as
    \ba
    \epsilon^2 \gamma_n^{(2)}+\epsilon^3 \gamma_n^{(3)}=\frac{c_1 \epsilon^2+c_2 \epsilon^3}{[\Delta_n(\Delta_n-1)]^{\Delta_{\phi^2}}}+\frac{\epsilon^2 d_1 (\gamma_{\phi^2}^{(1)}-1/3)+\epsilon^3 d_2(\gamma_{\phi^2}^{(2)}-\gamma_{\phi^2}^{(2),LRI})}{n^2}+\ldots
    \ea
    for some constants $c_{1,2}, d_{1,2}$. 
    Note that expanding in $\epsilon, n$ the first term leads to a $1/n^2$ fall off. We have rewritten the result suggestively following the expectations from \reef{assympgamman}.    
    The prescription is then that after subtracting off this term (which in perturbation theory `accidentally' gives a $1/n^2$ fall off), killing off the $1/n^2$ term above leads to the same constraint on the CFT data as we found from the super sum rule assuming cancellations of enhanced higher order contributions.

    \subsection{Long range $O(N)$}
    \label{subsec: O(N)}
    Now we turn to long-range scalar theories with quartic interactions in the presence of an $O(N)$ global symmetry. Let us first setup our conventions. Four-point functions of vectors fields are written as
    \begin{multline}
    \langle \phi_l| \phi_k(1)\phi_j(z)|\phi_i\rangle:=\mathcal G_{ij,kl}(z)\\
    =\delta_{ij} \delta_{kl}\, \mathcal G^{(S)}(z)+\frac 12 \delta_{i\{l}\delta_{k\}j}\, \mathcal G^{(T)}(z)+\frac 12 \delta_{i[l}\delta_{k]j} \, \mathcal G^{(A)}(z)\,.
    \end{multline}
    Crossing symmetry is now implemented by sum rules
    \ba
    \sum_{\mathbf b,\Delta} a_{\Delta}^{\mbf b} \beta_n^{\mbf a|\mbf b}(\Delta)=0\,, \quad \sum_{\mathbf b,\Delta} a_{\Delta}^{\mbf b} \alpha_n^{\mbf a|\mbf b}(\Delta)=0\,, \qquad \mbf a,\mbf b \in \{S,T,A\}
    \ea
    where the functionals are dual to a bosonic GFF $O(N)$ solution, which has spectrum
    \ba
    \Delta^{{\tt gff},S}_n=\Delta^{{\tt gff},T}_n=2\Df+2n\,, \quad\Delta^{{\tt gff},A}_n=1+2\Df+2n
    \ea
    in the appropriate channels with OPE coefficients $a_{n}^{{\tt gff} |\mbf a}=a_{\Delta}^{\tt gff} (1/N,1,1)^{|\mbf a}$. The functionals satisfy duality properties
    \ba
    \beta^{\mbf a|\mbf b}_n(\Delta_m^{{\tt gff},\mbf b})&=0\,,& \quad \partial_{\Delta}\beta^{\mbf a|\mbf b}_n(\Delta_m^{{\tt gff},\mbf b})&=\delta^{\mbf a,\mbf b} \delta_{n,m}\\
    \alpha^{\mbf a|\mbf b}_n(\Delta_m^{{\tt gff},\mbf b})&=\delta^{\mbf a,\mbf b} \delta_{n,m}\,,& \quad \partial_{\Delta}\alpha^{\mbf a|\mbf b}_n(\Delta_m^{{\tt gff},\mbf b})&=0\,.
    \ea
    More precisely, there are two absent functionals:
    \ba
    \beta_0^{S}=\beta_0^{T}=0
    \ea
    and as such the duality conditions on the righthand side are not correct for $\mathbf b=S,T$ and $m=0$. The reason for these absent functionals is again related to the existence of two relevant deformations arising from contact terms in AdS$_2$,
    \ba
    \int_{\mbox{\tiny AdS}}\left[ g_4^{(1)} (\Phi^2)^2 +g_4^{(2)} (\Phi_i \overset{\leftrightarrow}{\nabla} \Phi_j)^2\right]
    \ea
    In the flat space limit these lead to scattering amplitudes which can now be $O(s)$. If we want to fully span the space of deformations which in the flat space limit are not $o(1)$, we must also include the possibility of bulk exchange diagrams. Indeed, while these lead to scattering amplitudes which are $O(1/s^2)$ in the absence of $O(N)$ symmetry this is no longer the case where this symmetry is present.
    For instance the exchange of a state in the $A$ sector can arise from a vertex
    \ba
    \int_{\mbox{\tiny AdS}}
    g_e (\Phi_i \overset{\leftrightarrow}{\nabla} \Phi_j)\Psi^{[ij]}\,.
    \ea 
    and leads to a scattering amplitude which is $O(s)$. 
    
    Consider first perturbatively deforming the $O(N)$ GFF by one of these interactions. We write the corresponding changes in anomalous dimensions as  $(\gamma_{\partial^2 \Phi^4})^{\mbf a}_n\,, 
     (\gamma_{\Phi^4})^{\mbf a}_n, (\gamma_{\tt exc})_n^{\mbf a}$. Note that from the flat space limit we can find
     \ba
     (\gamma_{\partial^2 \Phi^4})^{\mbf a}_n=O(n^0)\,, \quad  (\gamma_{\Phi^4})^{\mbf a}_n=O(n^{-2})\,, \quad (\gamma_{{\tt exc}})_n^{\mbf a}=O(n^0)
     \ea
     Since these deformations can mix with each other we need to choose an appropriate basis. Our choice is to a) choose for the exchange a state in the $A$ channel with dimension $\Delta=2\Df-1$, and b) demand
     \ba
      (\gamma_{\partial^2 \Phi^4})^{\mbf a}_0=\left(\tfrac{N-1}{3},-\tfrac 1{3},-\tfrac{1}{(1+2\Df)}\right)\,,
      \quad
      (\gamma_{\Phi^4})^{\mbf a}_0=\left(\tfrac{N+2}6,\tfrac 13,0\right)\,, \quad
      (\gamma_{{\tt exc}})_0^{\mbf a}=\left(0,0,\tfrac{1-\Delta_\phi^2}{2(1+2 \Delta_\phi)}\right)
      \ea
      These choices completely fix the basis.
          Closed form expressions for the contact term anomalous dimensions are given explicitly in appendix \ref{app: anomalous dimension}. As for the exchange, we have by definition
      \ba
      (\gamma_{{\tt exc}})_n^{\mbf a}=-\frac{\beta_n^{\mbf a|A}(2\Df-1)}{a_{n}^{\tt gff}}\,.
      \ea
        
     Consider now a general correlator which in the UV is close to GFF. By acting with the $\beta_n^{\mbf a}$ sum rules and taking the large $n$ limit as we did before we find
    \begin{multline}
    \gamma_n^{\mbf a}\underset{n\gg 1}=(\gamma_{\partial^2 \Phi^4})^{\mbf a}_n \sum_{\mbf b,\Delta} a^{\mbf b}_{\Delta} \tilde \omega_1^{|\mbf b}(\Delta)+(\gamma_{\Phi^4})^{\mbf a}_n
    \sum_{\mbf b,\Delta}\, a^{\mbf b}_{\Delta} \tilde \omega_2^{|\mbf b}\\
    +(\gamma_{\tt exc})_n^{\mbf a} \sum_{\mbf b,\Delta}\, a^{\mbf b}_{\Delta} \tilde \omega_3^{|\mbf b}+(O(n^{-6})+{\tt n.a.})
    \end{multline}
    The {\tt n.a.}terms correspond to non-analytic corrections in $n$ such as $n^{-2\Delta}$. Let us discuss the functionals appearing above. As in the basis of $O(N)$ functionals we were missing $\beta_0^{S,T}$, above we certainly expect to see tilded versions of these functionals, $\tilde \beta_0^{S,T}$. These arise by relaxing the requirements on the fall-off conditions of functionals. However, when we do this, we get not two but actually three new functionals. The third functional has double zeros everywhere on the GFF spectrum. It is convenient to think of it as $\tilde \alpha_{-1}^A$. Concretely the duality conditions on these functionals are
    \ba
    \partial_{\Delta}\tilde \beta_0^{S|S}(\Delta^{{\tt gff},S}_m)=\partial_{\Delta}\tilde \beta_0^{T|T}(\Delta_m^{{\tt gff},T})=\delta_{m,0}\,,\\
    \tilde \beta_0^{S,T|A}(\Delta^{{\tt gff},A}_{-1})=0\,, \quad \tilde \alpha_{-1}^{A|A}(\Delta^{{\tt gff},A}_{-1})=1   
    \ea
    with $\Delta^A_{-1}=2\Df-1$, 
    and double zeros everywhere else on the GFF spectrum. With this choice of super functional basis we have the identifications:
    \ba
    \tilde \omega_1&=\tilde \beta_0^S-\tfrac{N+2}{2N} \tilde \beta_0^T\\
    \tilde \omega_2&= \tilde \beta_0^S+\tfrac{N-1}N \tilde \beta_0^T\\
    \tilde \omega_3&=\tilde \alpha_{-1}^A
    \ea
    Note in particular that the second functional above is the same as $\tilde \beta_0$ of the previous subsection as applied to $\mathcal G_{iiii}$,
    \ba
    \tilde \omega_2^{|\mbf a}=\left(1,\tfrac{N-1}N,0\right)^{\mbf a} \tilde \beta_0
    \ea

   In appendix \ref{App: Regge limit}, we explain how to obtain the Regge limit of Witten diagrams, and in appendix \ref{App: Regge super basis}, we provide a detailed prescription for constructing the Regge-superbounded basis. Let us now investigate whether the super sum rules for $\tilde \omega_{1,2,3}$ are satisfied by the long range $O(N)$ model. We begin with $\tilde \omega_2$, for which the computation is very similar to that of section~\ref{sec:lri}. Rescaling the functional by $\Df-\frac 14$ the sum rule is
    \ba
    \sum_{n,\mbf a} a_{n}^{\mbf a} \epsilon \tilde \omega_2^{|\mbf a}(\Delta_n^{\mbf a})=0.
    \ea
    To leading order we set $\Delta_n^{\mbf a}=\Delta_m^{{\tt gff},\bf a}+\epsilon \gamma_n^{(1)|\mbf a}$
    and expand the above for small $\epsilon$. Then we find
	\begin{multline}
	-\left[\gamma_0^{(1)|S}+(N-1)\,\gamma_0^{(1)|T} \right]\, \partial_{\Df} \partial_{\Delta}\hat \beta_0(2\Df)\bigg|_{\Df=\frac 14}+\\
    2 \left[\left(\gamma_0^{(1)|S}\right)^2+(N-1)\,\left(\gamma_0^{(1)|T}\right)^2 \right] \partial_{\Delta}^2\hat \beta_0(2\Df)\bigg|_{\Df=\frac 14}=0
	\end{multline}
    which simplifies to
    \ba
    -\gamma_0^{(1)|S}-(N-1) \gamma_0^{(1)|T}+3 \left[\left(\gamma_0^{(1)|S}\right)^2+(N-1)\left(\gamma_0^{(1)|T}\right)^2\right]=0 
    \ea
    Now let us consider $\tilde \omega_1$ which again we rescale by $\epsilon$. The sum rule now gives the simpler
    \ba
    \gamma_0^{(1)|S}-\frac{(N-2)}2 \gamma_0^{(1)|T}=0
    \ea
    as the $\partial^2_{\Delta}$ cancel out.
    Solving both constraints we find
    \ba
    \gamma_0^{(1)|S}=\frac{N+2}{N+8}\,, \quad \gamma_0^{(1)|T}=\frac{2}{N+8}
    \ea
    The first result is in precise agreement with \cite{Behan:2023ile}. Both happen to match the leading order results the $O(N)$ Wilson-Fisher model. We now proceed to the calculation at the next order. For this purpose, we also
require the first-order correction to the OPE coefficient, which is obtained by
solving the bounded functional $\alpha_0$,
\begin{equation}
    a^{(1)|S}_0=-\frac{2 (N+2) (\pi +4 \log (2))}{N (N+8)},\qquad
    a^{(1)|T}_0=-\frac{4 (\pi +4 \log (2))}{N+8}.
\end{equation}
Using this result, the constraint arising from $\tilde{\omega}_1$ takes the form
\begin{equation}
    \frac{2 \gamma^{(2)|S}_0}{N}
    -\frac{\gamma^{(2)|T}_0 (N+2)}{N}
    -\frac{\pi (N+2)}{(N+8)^2}
    -\frac{4 (N+2) \log (2)}{(N+8)^2}
    =0.
\end{equation}
Similarly, $\tilde{\omega}_2$ leads to the constraint
\begin{equation}
    - \frac{10 N+8}{N (N+8)}\gamma^{(2)|S}_0
    +\frac{2 N^2-10 N+8}{N (N+8)}\gamma^{(2)|T}_0
    +\frac{2 (\pi +4 \log (2))}{N+8}
    =0.
\end{equation}
Together, these equations uniquely determine the anomalous dimensions at
$O(\epsilon^2)$,
\begin{equation}
    \gamma^{(2)|S}_0=
    \frac{\left(7 N^2+34 N+40\right) (\pi +4 \log (2))}{(N+8)^3},
    \qquad
    \gamma^{(2)|T}_0=
    -\frac{\left(N^2-6 N-40\right) (\pi +4 \log (2))}{(N+8)^3}.
\end{equation}
The dimension of the singlet operator is consistent with existing results,
whereas for the traceless symmetric operator our result constitutes a
prediction.

Let us now turn to the rescaled third functional, $\epsilon\,\tilde{\omega}_3$.
\emph{A priori}, it is not obvious that this sum rule can be satisfied while
keeping the results derived from the other two functionals unchanged.
Remarkably, however, we find that this is indeed the case, due to two
nontrivial mechanisms that provide a strong consistency check of our proposal.
First, by construction, almost all double-trace operators lie on double zeros of the functionals implying that their contributions are proportional to
$a^{(0)|i}_n (\gamma^{(1)|i}_n)^2$ in all sectors $i=S,T,A$.
Since $\gamma^{(1)|i}_n=0$ for $n>0$ in the singlet and traceless symmetric
sectors, and $\gamma^{(1)|A}_n=0$ for all $n$ in the antisymmetric sector,
these operators begin contributing only at $O(\epsilon^4)$. On the other hand, the leading $\phi^2$ operator in the singlet and traceless
symmetric channels ($n=0$) could, in principle, contribute at $O(\epsilon^2)$.
However, we find that for those operators the functional develops a quadruple zero, and as a
result, these contributions are also pushed to $O(\epsilon^4)$.

\subsection{Long range Lee-Yang}
In this section we investigate our sum rule applied to the long-range Lee-Yang model.  It is defined as the critical point of the theory
    \ba
    S=\int_{\mbox{\tiny AdS}} \left[ (\nabla \Phi)^2+m^2 \Phi^2\right]+i g \int_{\partial \mbox{\tiny AdS}}\, \phi^3
    \ea
    For $\Df<\frac 13$ the operator $\phi^3$ is relevant and the theory flows to a non-trivial, non-unitary (but PT-symmetric), fixed point. 
    Setting $\epsilon=(1-3\Df)\ll 1$ it is possible to compute the data of this CFT perturbatively in $g^*\sim \sqrt{\epsilon}$. Note that the theory breaks $Z_2$ so that $\phi$ can appear in its own OPE:
    \ba
    \phi \times \phi=1+ \phi+\phi^2+\ldots
    \ea
    One important comment is that for this theory we have the relation
    \ba
    \Delta_{\phi^2}=1-\Delta_{\phi}=2\Df+\epsilon
    \ea
    which follows from the free bulk equation of motion.
    Furthermore PT symmetry, or equivalently, symmetry under $i\to -i$ and $\phi \to -\phi$ tells us that the OPE coefficient $\lambda_{\phi \phi \phi}$ must be purely imaginary. 
    
    Since the dimension of $\phi^2$ is fixed by the above our goal is to compute the OPE coefficient $\lambda_{\phi \phi \phi}$ from the super sum rule as applied to the  $\langle \phi \phi \phi \phi\rangle$ correlator. 
    
    The correlator is real and hence admits an expansion in integer powers of $\epsilon$. We write
    \ba
    \lambda_{\phi \phi \phi}^2=\epsilon a_{\phi}^{(1)}+\epsilon^2 a_{\phi}^{(2)}+\ldots
    \ea
    Let us begin with the $O(\epsilon)$ computation. At this order the sum rule receives contributions from $\phi$ and $\phi^2$ only, and we get simply
    \ba
    a_{\phi}^{(1)}=-\frac{a^{\tt gff}_0}{\tilde \beta_0(\Df)}\bigg|_{\Df=\frac 13}=-\frac{3\sqrt{3}\, \Gamma(\tfrac 13)^3}{4\pi}
    \ea
    We can also compute anomalous dimension of double trace operators. Writing $\Delta_n=2\Df+2n+\epsilon \gamma_n^{(1)}$ we get
    \ba
    \gamma_n^{(1)}=-\frac{a_{\phi}^{(1)}}{a_n^{\tt gff}}\, \beta_n(\Df)\bigg|_{\Df=\frac 13}+\frac{a_0^{\tt gff}}{a_n^{\tt gff}}\, e_n \bigg|_{\Df=\frac 13}
    \ea
    where we recall the $e_n$ appear in \reef{eq:crossing}. We were not able to find a closed form expression for the above, but we could check the leading asymptotics
    \ba
    \gamma_n^{(1)}\underset{n\gg 1}\sim \frac{1}{\Gamma(\tfrac 13)}\, \frac{1}{[\Delta_n^{\tt gff}(\Delta_n^{\tt gff}-1)]^{\frac 13}}+O(n^{-8/3})
    \ea
    consistently with \reef{assympgamman}. Finally we will also need the correction to the $\phi^2$ OPE coefficient,
    \ba
    a_0=a_0^{\tt gff}+\epsilon a_0^{(1)}
    \ea
    given by
    \ba
    a_0^{(1)}=\left[-a_{\phi}^{(1)} \alpha_0(\Df)+f_0\right]\Bigg|_{\Df=\frac 13}= -\frac{4 \pi }{\sqrt{3}}-6 \log (3)+\frac{2 \pi ^2}{\Gamma \left(\frac{1}{3}\right)^3}
    \ea

    Now let us move on to the computation at the next order. The main difficulty is the tower of double traces which give a divergent contribution to the super sum rule. However, unlike the long range Ising case, this does not lead to an enhancement in the order of the perturbative expansion. We have
    \ba
    \sum_{n} a_n^{\tt gff} \tilde \beta_0(\Delta_n)\underset{n\gg 1}{\sim}  \sum_n c\, n^{-\frac 13}\,, \quad \mbox{with} \quad c:=\frac{4 \pi ^2 \epsilon^2}{\Gamma \left(\frac{1}{3}\right)
   \Gamma \left(\frac{5}{6}\right)^4}
    \ea
    Thus we define
    \ba
    \tilde\beta_0[{\tt tower}]:=\sum_{n=1}^{\infty} a_n^{\tt gff} \tilde \beta_0(\Delta_n) \quad \Longrightarrow \quad \sum_{n=1}^{\infty} \left[a_n^{\tt gff} \tilde \beta_0(\Delta_n)-c\, n^{-\frac 13}\right]+c\,\zeta(1/3)
    \ea
    The equation we must solve is now
    \ba
    \underbrace{a_{\phi}^{(2)} \tilde \beta_0(\Df)-\frac 13 a_\phi^{(1)}\partial_{\Df} \tilde \beta_0(\Df)}_{\tilde \beta_0[\phi]}+\underbrace{a_0^{(1)}+a_0^{\tt gff} \partial_{\Delta}^2\tilde\beta_0(2\Df)}_{\tilde \beta_0[\phi^2]}+\tilde \beta_0[{\tt tower}]=0
    \ea
    evaluated at $\Df=\frac 13$.
    While it is difficult to evaluate the various terms above analytically, nevertheless we have checked numerically to high accuracy that the result is:
    \ba
    a_{\phi}^{(2)}=
    -\frac{3 \Gamma \left(\frac{1}{3}\right)^3 \left(8 \pi ^2+4 \sqrt{3} \pi  \log (27)+9 \Gamma \left(\frac{1}{3}\right)^3\right)}{32 \pi ^2}
    \ea
    This result, together with $a_{\phi}^{(1)}$, agree with those computed explicitly in the appendix \ref{app: Lee-Yang} using the lagrangian definition of the theory. Similarly to the Long Range Ising model, we have also checked that the same results follow if, rather than using the $\tilde \beta_0$ sum rule, we compute all perturbative quantities using crossing and demand that the $n^{-2}$ term in $\gamma_n$ is absent.

    \section{Numerical applications?}
    \subsection{Applicability of sum rule}
In this section we begin exploring implications of the $\tilde \beta_0$ sum rule for numerical bootstrap bounds. The main difficulty in this study is the fact the sum rule does not hold for a generic CFT correlator: as we've seen, suitable constraints on the Regge limit are a priori necessary for the sum rule to make sense. While it can be possible to relax these constraints, to do this we have to regularize the sum rule, which requires additional information on the Regge limit. The previous section demonstrated that a suitable regularization can indeed lead to meaningful results. 
Thus, when discussing the $\tilde \beta_0$ sum rule there are several possibilities:

\begin{enumerate}
\item We take the sum rule at face value. The resulting constraints hold for CFT correlators sufficiently bounded in the Regge limit, as discussed in section \ref{sec:transparency}.
\item We allow for regularization of the sum rule to account for slow fall off of anomalous dimensions of GFF like states. 
\item We allow for additional contributions requiring regularization, namely towers of multi-trace operators with OPE coefficients decaying insufficiently fast.
\end{enumerate}

Let us discuss these in turn. In the first case the class of CFT correlators are GFF like in the UV, i.e. with the OPE density peaked around dimensions $\Delta_n=2\Df+2n$. The average squared anomalous dimension of states localized near these values has to decay at least as $1/n^2$, as should contributions to the OPE away from them (relative to the GFF OPE density).

In the second case, we relax this slightly by allowing slower fall off of the anomalous dimension squared. Concretely we assume
\ba
\langle \gamma_n\rangle^2=\langle \gamma_n^2\rangle=C n^{-\alpha} + o(n^{-2})
\ea
In this case, on the one hand the sum rule can be regulated by setting
\ba
\sum_{n=1}^\infty n^{1-\alpha}:=\zeta(1-\alpha)
\ea
On the other, both $\alpha$ and $C$ are actually fixed by
crossing symmetry in terms of the CFT data of the lowest dimension operator: using \reef{assympgamman} we get
\ba
C\propto \lambda^2_{\phi \phi \cO} \,, \quad \alpha=4\Delta_{\cO}
\ea
Incidentally note that these results mean that CFTs with operators of dimension smaller than $1/2$ cannot be considered in possibility one.

Finally in the last case we also relax the constraint on contributions of the OPE density away from GFF values. This is of course so broad that a priori any kind of CFT correlator could be considered. But without knowing what exactly is happening in the UV we cannot do this regularization. Concretely we are allowing
\ba
\langle \gamma_n\rangle< \langle \gamma_n^2\rangle=C_1 n^{-\alpha_1}+C_2 n^{-\alpha_2}+\ldots+o(n^{-2})\,, \qquad 0\leq \alpha_i<2
\ea
Each `trajectory' above can be regulated separately but we cannot now use (single correlator) crossing to help us determine the various constants. As such they must be given to us as inputs characterizing the class of CFT correlators under consideration. 

\subsection{Applications}

\begin{figure}
    \centering
    \includegraphics[width=0.8\linewidth]{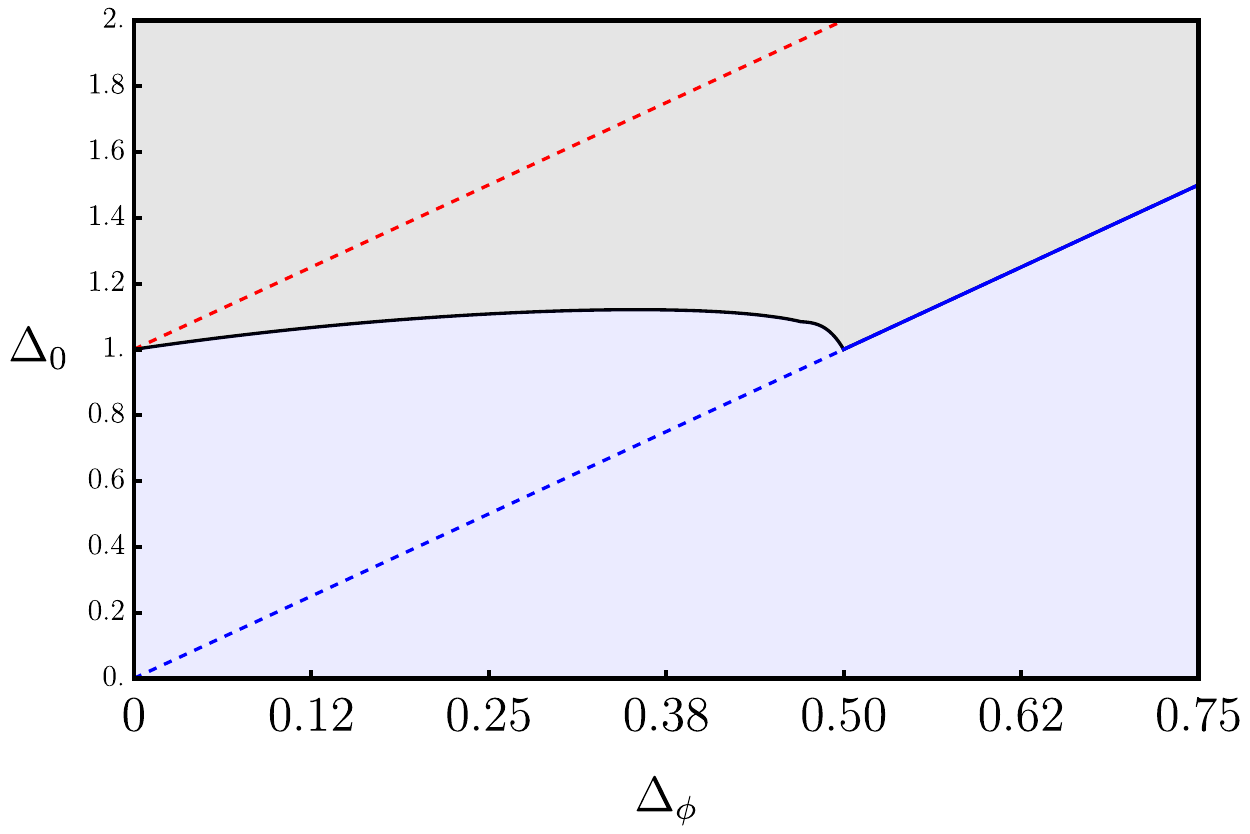}
    \caption{Bound on the gap above the identity in the $\phi \phi$ OPE, when imposing the $\tilde \beta_0$ sum rule. For $\Df\geq 1/2$, the bound is the generalized free boson $\D_0=2\Df$. The bound becomes non trivial for $0\leq \Df<1/2$. The red dashed curve is the maximum gap without imposing the $\beta_0$ sum rule, corresponding to the generalized free fermion at $\D_0=2\Df+1$.}
    \label{fig:gapmax}
\end{figure}

As our first application we consider the gap maximization problem, i.e. the problem of establishing an upper bound on the dimension $\Delta_0$ of the first operator in the $\phi\times \phi$ OPE. In $d=1$ it is well established that using crossing alone the optimal bound is given by $1+2\Df$ and saturated by the generalized free fermion correlator\cite{Gaiotto:2013nva, Mazac:2016qev}. Upon including $\tilde \beta_0$ we find the bound is dramatically improved, as shown in figure \ref{fig:gapmax}. 

The result can be understood straightforwardly from the plots of $\tilde \beta_0$ in figure \ref{fig: superbeta}. For $\Df\geq 1/2$ the functional is positive above $2\Df$ (and zero on the identity) so this automatically gives us a bound $\Delta_0\leq 2\Df$. The bound is optimal as it is saturated by the generalized free boson solution~\reef{gffcorr}, and the extremal functional for this problem is simply $\tilde \beta_0$. For $\Df\leq 1/2$ we lose positivity immediately above $2\Df$ because of the pole in $\tilde \beta_0$, which now gives a bound $\Delta_0\leq 1+{\tt small}$. Adding further functionals does not modify this significantly.\footnote{In particular the bound cannot be improved below $\Delta_0=1$, since ordinary crossing functionals are finite while $\tilde \beta_0$ has a pole.} The extremal solution corresponding to this bound is simple to understand: it is a particular member of the family of OPE max solutions to crossing \cite{Paulos:2019fkw}. This family is obtained by maximising the OPE coefficient of an operator of dimension $\Delta_0$, and morally corresponds to a $\Phi^4$ theory in AdS$_2$. For $\Df\leq 1/2$, it is possible to choose a particular $\Delta_0$ so that the $\tilde \beta_0$ sum rule is satisfied.%
\footnote{In perturbation theory these solutions have $g_4^{\mbox{\tiny eff}}=0$ due to a cancellation between one-loop and tree level diagrams.}

Let us now try to get a more interesting result. Inspired by the long-range Ising model, we allow for an operator below $\Delta=1$ and maximize the gap above it. In practice we ask for a functional satisfying:
\ba
\omega(\D_{\phi^2})&=\omega(\D_{\phi^4})=0\\
\partial_{\Delta}\omega(\D_{\phi^2})&=0\\
\omega(\D)&\geq 0\quad \mbox{for}\ \D\geq \D_{\phi^4} \text{ and } \D_-\leq \D<1
\ea
Finally we demand:
\ba
\D_{\phi^2}>\frac 12
\ea
the reasons for which will be explained below.
This establishes an upper bound on the dimension $\Delta_{\phi^4}$ and is saturated by a solution containing an operator at $\Delta_{\phi^2}$. Note that the double zero constraint at $\Delta_{\phi^2}$ is what determines its value. Equivalently we can also simply drop this constraint and scan over $\Delta_{\phi^2}>\frac 12$ to find the best possible bound.

\begin{figure}[t]
    \centering
        \includegraphics[width=0.95\linewidth]{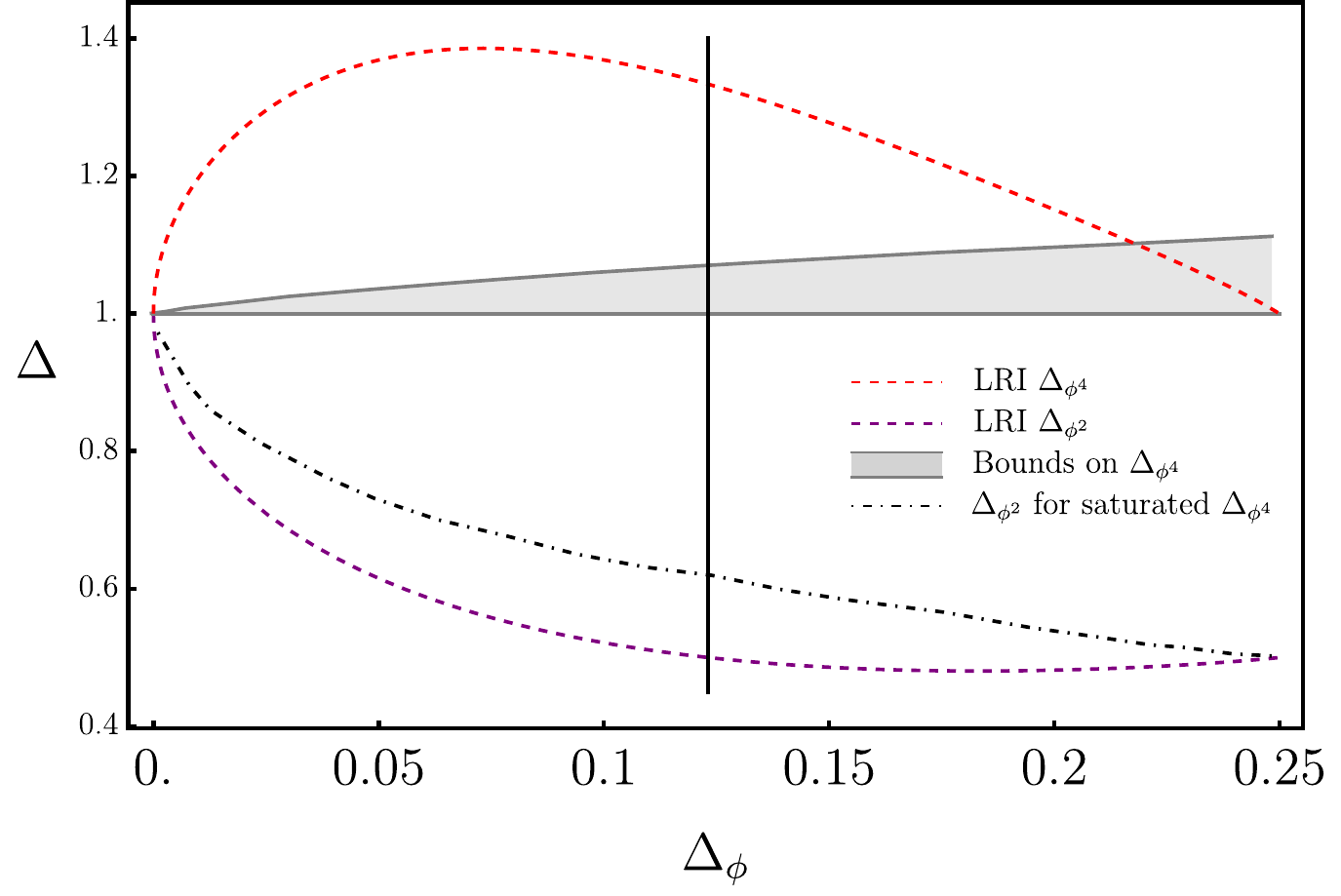} 
        \caption{Bounds on near marginal operators and comparison with Long range Ising perturbative data. The gray region is a bound on the dimension of an operator above 1, allowing operators with $\frac 12<\Delta<1$. When the bound is saturated one finds an operator in the extremal solution whose dimension is shown as the black dot dashed curve. For comparison, we show the dimensions of the operators $\phi^2$ and $\phi^4$ for the long range Ising model obtained from Pad\'e resummations of available perturbative data, plotted as the bottom and top red dashed curves respectively. The black vertical line divides the plot into two regions according to which the resummed $\Delta_{\phi^2}$ is larger or smaller than $1/2$; in the former our bound does not apply to LRI anyway.
    \label{fig:dphi4dphi2plot}
    }
\end{figure}

The result is shown in figure \ref{fig:dphi4dphi2plot}. The bound on the gap above 1 is only slightly above the previous bound on the overall gap. To each point is associated an extremal solution to crossing, whose first operator lies between 1/2 and 1 and is shown at the bottom of the same figure.
We compare these results with the available perturbative data for the long range Ising model \cite{Benedetti:2020rrq,Benedetti:2025nzp}. Near $\Df=\frac 14$, setting $\epsilon =1-4\Df$, we have 
\ba
\Delta_{\phi^2}&=\frac 12-\frac{\epsilon}6+\frac{\epsilon^2}{9}[\pi+\log(16)]+\frac{\epsilon^3}{81} \left(-\alpha_{I_4}+\pi ^2+(\pi +\log (4)) \log (64)\right)\\
\Delta_{\phi^4}&=1+\epsilon-\frac{2}{3} \epsilon ^2 (\pi +\log
   (16))+\frac{2}{27} \epsilon ^3 \left(\alpha_{I_4}+3 \pi  (3 \pi -8)+156 \log ^2(2)+78 \pi  \log (2)\right)
\ea
where $\alpha_{I_4}\sim 330$. For $\epsilon$ close to one we have instead
\ba
\Delta_\phi^2&=1-\sqrt{1-\epsilon}+\frac{1-\epsilon}8\\
\Delta_{\phi^4}&=1+\sqrt{1-\epsilon}+\frac{1-\epsilon}8
\ea
In the plot we show Pad\'e resummations of this perturbative data.
We see that the bound compares unfavourably with the resummed data. Firstly, focus on the left hand side of the plot, where $\D_{\phi^2}$ for LRI is larger than $\frac 12$ and hence satisfies our assumptions. In this region our bound seemingly rules out the LRI model. We conclude that our assumption of no subtractions, i.e. possibility 1 discussed above, must be incorrect (or of course that the sum rule does not apply to LRI non-perturbatively). We will comment on this further below. 

To the right of the plot our bound is not really valid for LRI, since we are in the region where $\Delta_{\phi^2}<\frac 12$. In that region we should really relax our assumption on $\Delta_{\phi^2}$ and recompute our bound. We have done just that, for simplicity setting its dimension to be that of LRI and regularising the contribution of double trace states, as discussed in possibility 2 above. In this case we find that the bound on $\phi^4$ is now drastically weakened, so much so that it is now indeed compatible with the LRI result, but boringly so: in particular as $\Df\to \frac 14$ the bound stays far above 1. The extremal solution turns out to be the OPE max solution for $\D_{\phi^2}$, so that the bound is roughly $2+2\Df$. To see why, note that after regularization the sum rule is something like
\ba
a_0 \tilde \beta_0(\Delta_{\phi^2})+\sum^{\Delta^*}_{\Delta>\Delta_{\phi^2}} a_{\Delta} \beta_0(\Delta)=a_0\, \frac{(\Delta^*)^{2-4\Delta_{\phi^2}}}{2-4\Delta_{\phi^2}}+a_0 \times (\tt finite, positive)
\ea
The large term on the right hand side cancels with the tail of the sum over double trace operators on the left. As for the remaining finite positive term on the right hand side we can match it by adding to the OPE max a solution an operator of dimension $\sim \Delta^*$ and OPE coefficient of order $a_{\Delta^*}^{\tt gff}/\Delta^*$. This is invisible to the remaining crossing sum rules, but not to $\tilde \beta_0$ due to its slower fall off at large $\Delta$.

Let us now comment on the need for additional subtractions. As in the simple case above, such subtractions amount to including extra positive contributions to the right hand side of the $\tilde \beta_0$ sum rule. This will accordingly relax the bound. That such contributions must be there does indeed seem evident from our perturbative analysis in section \ref{sec:lri}, which shows that $(\phi^4)_n$ type operators are present and require regulation. Further evidence for this is the proposal of \cite{Benedetti:2025nzp} for a weakly coupled description of LRI near $\Df=0$. In this proposal the correlator $\langle \phi \phi \phi \phi\rangle$ was computed to $O(\Delta_\phi^2)$ where it was found to contain a tower of operators which do not lie on the GFF dimension values. Unfortunately this means that, at least with a single correlator, we will not be able to get interesting bounds satisfied by LRI. Instead the mixed correlator system including $\phi,\phi^2$ appears to be the minimal such system leading to interesting results, as it would allow us to bootstrap a $(\phi^4)_n$ type tower.

    \section{Discussion}
In this paper, we motivated extra sum rules that may be imposed when we consider a free bulk theory in $AdS_2$. There are an infinite number of sum rules that can be imposed on the CFT data,  motivated by the large energy expansion of asymptotically S-matrix. We have explored the consequences of only the first such sum rules in perturbation theory and argued that it seemingly isolates the long-range theories. In particular, we have studied the long-range Ising model and its O(N) cousin, and also the non-unitary Lee-Yang model.  Then we have applied these sum rules to study the space of unitary solutions nonperturbatively. Due to the absence of multiparticle states, our bounds are not saturated by the LRI, but we see a huge reduction of space of solutions.

\subsubsection*{SRI end of 1D LRI}

In this paper we have tested super sum rules for long range models close to mean field theory. Recently, in~\cite{Benedetti:2024wgx}, a topological model coupled to a
compactified generalized free field was proposed to describe the long range Ising model close to the short range 1d Ising (topological) CFT. Let us briefly investigate the validity of the
$\tilde{\beta}_0$ sum rule for the perturbative CFT data obtained
in~\cite{Benedetti:2025nzp} around the short-range regime. 

The perturbative scheme in~\cite{Benedetti:2024wgx} is organized as a power series in
$\Df$. At leading order, the $\phi$ correlator is equal to unity. At $O(\delta)$,
two operators of dimension $1$ contribute, identified as $\phi^2$ and $\phi^4$. At $O(\delta^2)$,
new operators with dimensions $\Delta=n$, $n\geq 2$, appear with OPE
coefficients
\begin{equation}
   a^{(2)}_n=
   \frac{2^{3-2 n} \big((n-1) n+2\big) \Gamma (n-1)}
   {(n-1) n^2 \left(\frac{3}{2}\right)_{n-2}}.
\end{equation}

Let us consider the super sum rule suitably regulated as
$\hat{\beta}_0=\Delta_{\phi}^3 \tilde \beta_0$. Expanding to $O(\delta)$ we find that both operators of
dimension $1$ contribute with the same sign so that the sum rule is not satisfied. However, as is by now familiar, there is an enhancement arising
from the contribution of tower of operators that naively appears at $O(\delta^2)$. 

The contribution from operators with $\Delta=2n$ drops out, since they sit at
double zeros of the functional. On the other hand, for the remaining states we find
\begin{equation}
    a^{(2)}_n\,\hat{\beta}_0(2n+1)
    \sim \frac{4}{n^2}+4 n+\frac{8}{n}-2.
\end{equation}
Applying zeta-function regularization the $1/n$ term leads to an enhancement. Writing the OPE coefficients as
$a^{(2)}_n\sim n^{-1-c_0 \delta}$, cancellation of these terms with the contribution of two operators with dimension close to 1 requires
$c_0=-2$. It would be very interesting to verify
this prediction by computing higher-order CFT data in the future.

\subsubsection*{Long range Ising: general $d$}
Let us consider possible generalisations of our story to higher-d CFT. We begin by looking for special features in the perturbative data near the long-range Ising point. First, consider that in general dimension $d$, the anomalous dimensions induced by the $\Phi^4$ interaction take the form
\begin{equation}
 \gamma_{n>0,0}=
 \frac{2^{-2 n-1} n!
 \left(\frac{d}{2}\right)_n
 (\Delta_\phi)_n
 \left(2 \Delta_\phi-\frac{d}{2}\right)_n
 (-d+n+2 \Delta_\phi+1)_n}
 {\Gamma(n+1)^2
 \left(\Delta_\phi+\frac{1}{2}\right)_n
 \left(-\frac{d}{2}+\Delta_\phi+1\right)_n^2}
 \,\gamma_{0,0}.
\end{equation}
In the large-$n$ limit, this expression behaves as
\begin{equation}
   \gamma_{n,0}\propto \frac{n^{d-3}}{\Gamma(2\Df-d/2)}
\end{equation}
In particular at $\Delta_\phi=d/4$, where the long range Ising model lies, the anomalous dimensions vanish just as for $d=1$.

In one dimension we have extracted the super sum rule in two ways. Firstly by considering a general perturbation of a GFF in the UV and using crossing symmetry, under the guise of functionals, to read off the $1/n^2$ term. Here the analogous thing to do would be to extract the $n^{d-3}$ piece. Alternatively, the super sum rule arises from a functional applicable to a super Regge bounded correlator. Such functionals have been constructed \cite{Mazac:2019shk} but it is not clear which ones should be considered to match our $d=1$ computation. 

Instead we have noticed the following observation. Functional actions can be extracted by considering crossing symmetric sums of Witten diagram exchanges in AdS$_{d+1}$ together with suitable choices of contact terms. These are fixed by writing suitable dispersion relations in Mellin space \cite{Gopakumar:2021dvg, Bhat:2025zex}.
We have found that, if we ignore contributions from contact terms (which is not motivated by any known prescription for constructing a good crossing symmetric functional basis), then we can extract a `functional' $\tilde \beta_{0,0}$ (the coefficient of the $n=0,\ell=0$ double trace operator in the block expansion), whose sum rule gives perturbatively, as we did in $d=1$:
\begin{equation}
\label{eq:eps}
\begin{split}
    & \gamma^1_{0,0}=\frac{1}{3},\\
    & \gamma^2_{0,0}=\frac{1}{9} \left(\psi \left(\frac{d}{2}\right)-2 \psi \left(\frac{d}{4}\right)+\psi (d)\right).
    \end{split}
\end{equation}
These precisely match the LRI data in general $d$ \cite{Fisher:1972zz}.

Qualitatively, the key difference between superbounded and bounded bases is that scalar operators contribute to locality constraints in the superbounded sum rules. Indeed, it is precisely this feature that gives rise to the additional contact contribution in the Polyakov sum rules. Since we do not know the coefficient of the $n^{d-3}$ asymptotics of the higher-dimensional functionals, we are unable to draw any definitive conclusions at this stage. Nevertheless, it is plausible that the sum rule associated with the high-energy expansion effectively eliminates this contact contribution in the scalar sector.
\\
\\
Finally, we outline some directions for future work. One immediate goal is to
perform a numerical mixed-correlator bootstrap with these additional sum rules
imposed. We expect that the resulting bounds will be saturated by LRI solutions.
So far, we have imposed only the $\Sigma_0$ sum rule arising from the free bulk
description. It would be interesting to investigate how the remaining sum rules
further constrain the space of solutions. It is tempting to speculate that one
might observe saturation of the numerical bounds by LRI solutions even in a
single-correlator bootstrap once these additional constraints are imposed, since
they would force the spectrum to include multiparticle states, which are
necessary for saturation.

In addition, our strategy for computing the OPE coefficients of four-particle
states built from the fundamental fields can be extended to other classes of
operators, allowing for the computation of mixed correlators involving
composite operators in perturbation theory \cite{Ghosh:2023wjn, Bertucci:2022ptt}.

Another interesting application of our super sum rules concerns quantum field
theories in $AdS_2$ \cite{Antunes:2024hrt,Copetti:2023sya, Carmi:2018qzm, Hogervorst:2021spa}. In particular, if one wishes to bootstrap a specific theory,
such as a model with a $\Phi^4$ interaction in $AdS_2$, analogous sum rules can
be employed to restrict the space of allowed solutions. It was shown
in~\cite{Paulos:2019fkw} that, at weak coupling, the bound on the OPE coefficient
of the leading operator is saturated by this theory. Owing to the absence of
multiparticle states, this bound differs from that of the true $\Phi^4$ theory
in the bulk at large coupling. A mixed-correlator study was performed in~\cite{Ghosh:2023wjn}.
However, even after imposing appropriate gaps, the resulting allowed islands
remain relatively large. Therefore, it would be interesting to incorporate the
additional sum rules for mixed correlators by relating their Regge limit in the presence of $\Phi^4$ interaction and investigating
whether they lead to sharper allowed regions. We leave these and other
interesting applications for future work.

\subsection*{Acknowledgements}
It is a pleasure to acknowledge discussions with Ant\'onio Antunes, Connor Behan, Ian Moult, and Philine van Vliet. This work was co-funded by the European Union (ERC, FUNBOOTS, project number 101043588). Views and opinions expressed are however those of the author(s) only and do not necessarily reflect those of the European Union or the European Research Council. Neither the European Union nor the granting authority can be held responsible for them. Z.Z. is supported by Simons Foundation grant \#994308 for the Simons Collaboration on Confinement and QCD Strings. KG is supported by the Royal Society under grant RF \textbackslash ERE\textbackslash    231142.

\newpage
\appendix  
\section{The $\tilde \beta_0$ functional}
\label{app:beta0}
In this appendix we give more information on the $\tilde \beta_0$ functional. The functional action can be written as
\ba
\tilde \beta_0(\Delta)=\int_{\frac 12}^{\frac 12+i\infty} \ud z f(z) F_{\Delta}(z)+\int_{\frac 12}^1 \ud z g(z) F_{\Delta}(z)
\ea
with $F_{\Delta}(z)=G_{\Delta}(z)-G_{\Delta}(1-z)$ and
\ba
g(z)=(1-z)^{2\Df-2} f(\tfrac{1}{1-z})\,, \qquad f(z)\underset{z\to \infty}=\frac{(2\Df-1)^2}{2\pi^2\kappa_{\Df}}
\ea
For instance, for $\Df=1$ we have
\ba
f(z)=\frac{2}{\pi^2}\frac{1+z(z-1)}{z(z-1)}
\ea
It is possible to derive an explicit expression for $f(z)$ for any $\Df$ but this will not be useful for us. Instead let us give the functional action:
\begin{multline}
\tilde \beta_0(\Delta)=\frac{\pi  2^{-2 \Delta _{\phi }-2 \Delta +3} \left(\Delta -2 \Delta _{\phi }\right) \Delta _{\phi } \left(2 \Delta _{\phi }+\Delta -1\right)
   \Gamma (2 \Delta ) \Gamma \left(2 \Delta _{\phi }-\frac{1}{2}\right) \Gamma \left(\Delta _{\phi }\right){}^3}{(\Delta -1) \Delta  \Gamma
   \left(\frac{\Delta }{2}\right)^4 \Gamma \left(\Delta _{\phi }-\frac{1}{2}\right) \Gamma \left(-\frac{\Delta }{2}+\Delta _{\phi }+1\right){}^2
   \Gamma \left(\frac{\Delta +1}{2}+\Delta _{\phi }\right){}^2}+\\
   {\tt PFQ}(\Delta,\Df)\,\times 
   \frac{2^{\Delta } \Delta   \Gamma \left(\frac{\Delta +1}{2}\right)
   \Gamma \left(2 \Delta _{\phi }-\frac{1}{2}\right) \Gamma \left(\Delta +2 \Delta _{\phi }-\frac{1}{2}\right) \Gamma \left(\Delta _{\phi
   }\right){}^4}{\sqrt{\pi } \Gamma \left(\frac{\Delta }{2}+1\right) \Gamma \left(\Delta _{\phi }-\frac{\Delta }{2}\right){}^2 \Gamma
   \left(\frac{\Delta }{2}+\Delta _{\phi }\right){}^2 \Gamma \left(\frac{1}{2} \left(\Delta +4 \Delta _{\phi }-1\right)\right){}^2}
\end{multline}
with
\ba
{\tt PFQ}(\Delta,\Df)=\pFq{7}{6}
{\frac{\Delta }{2},\frac{\Delta }{2},\Delta _{\phi }+\frac{\Delta -1}{2},\Delta _{\phi }+\frac{\Delta -1}{2},\Delta _{\phi }+\frac{\Delta
   }{2}+\frac{1}{4},2 \Delta _{\phi }-1,2 \Delta _{\phi }+\Delta -\frac{3}{2}}
{\Delta +\frac{1}{2},\Delta _{\phi }+\frac{\Delta }{2}-\frac{3}{4},\Delta _{\phi }+\frac{\Delta }{2},\Delta _{\phi }+\frac{\Delta
   }{2},\frac{1}{2} \left(4 \Delta _{\phi }+\Delta -1\right),\frac{1}{2} \left(4 \Delta _{\phi }+\Delta -1\right)}
{1}
\ea

Swapping is the property
\ba
\tilde \beta_0[\mathcal G(z)-\mathcal G(1-z)]=\sum_{\Delta} a_{\Delta} \tilde \beta_0(\Delta)
\ea
The behaviour near $z=\infty$ implies that $\tilde \beta_0$ satisfies swapping only for correlators~$\mathcal G(z)\underset{z\to \infty}=O(z^{-1-\epsilon})$ for $\epsilon>0$. In particular it is easy to see that this excludes ordinary correlators with identity or indeed containing any operators dimension $\Delta_0\leq 1$, since it is only possible to bound the correlator as $z^{-\Delta_0}$ in the Regge limit. 

Suppose now that we have a crossing symmetric function  satisfying
\ba
\widehat{\mathcal G}(z)\underset{z\to \infty}=i\pi c/z+\,,\qquad \mbox{Im} z>0\ldots
\ea
and let us set $\lim_{z\to \infty} f(z)=\kappa$.
Then it is easy to find, by rotating contours as in \cite{Mazac:2018mdx},
\ba
0=\tilde \beta_0[\widehat{\mathcal G}(z)-\widehat{\mathcal G}(1-z)]=\int_0^1 g(z)\mbox{dDisc}\, \widehat{ \mathcal G}(z)-c \frac{(\Df-1/2)^2}{\kappa_{\Df}}
\ea
Choosing $\widehat{ \mathcal G}=\mathcal G(z)-\sum_{\Delta\leq 1} a_{\Delta} \mathcal P_{\Delta}(z)$ with $\mathcal P_{\Delta}$ a super-Regge bounded Polyakov block \cite{Mazac:2019shk,Paulos:2020zxx} the above becomes precisely
\ba
\kappa_{\Df}\sum_{\Delta\geq 0} a_{\Delta} \tilde \beta_0(\Delta)=c \left(\Df-\tfrac 12\right)^2 \label{beta0largez}
\ea
so that the supersum rule reads off the $1/z$ term in the correlator. Note that summing blocks with small anomalous dimensions $\gamma_n\sim 1/n^2$ gives
\ba
n^2 \gamma_n \underset{n\to \infty}= b\quad \Longrightarrow \quad \mathcal G(z)\underset{z\to \infty}=\frac{i \pi b/(\Df-\frac 12)^2}{z}+\ldots
\ea
so that the \reef{beta0largez} above is precisely consistent with \reef{assympgamman}.

\section{Determination of proportionality constant}
\label{app:constant}
In this appendix we determine the precise RHS of the super sum rule in terms of the bulk coupling. We work with the bulk action
\ba
\int_{\mbox{\tiny AdS}} \frac{\ud y\ud x}{y^2}\,\left[\frac 12 (\nabla \Phi)^2+\frac 12 m^2 R_{\mbox{\tiny AdS}}^2  \Phi^2 +\frac{g_4}{4!} R_{\mbox{\tiny AdS}}^2\, \Phi^4\right]
\ea
We set
\ba
\Phi(y,x)\underset{y\to 0}=\sqrt{C_{\Df}}\,y^{\Df} \phi(x)\,, \quad C_{\Df}:=\frac{\Gamma(\Df)}{2\sqrt{\pi}\Gamma(\frac 12+\Df)}
\ea
so that the boundary field $\phi$ has the usual CFT two point function with unit coefficient. Setting $\mathcal G(z)=\langle \phi|\phi(1)\phi(z)|\phi\rangle$,  to leading order the effect of AdS contact interaction is given by
\ba
\delta \mathcal G(z)=g_4 C_{\Df}^2 \, D_{\Df}(z)
\ea
with $D_{\Df}(z)$ the D-function, defined with normalisation
\ba
D_{\Df}(z)=\int \frac{\ud x \ud y}{y^2}\, \left[\frac{y^4}{(y^2+x^2)(y^2+(x-1)^2)(y^2+(x-z)^2)}\right]^{\Df}
\ea
In particular this gives
\ba
\gamma_{\phi^2}=\frac{g_4}{16\pi \kappa_{\Df}}\,, \qquad \kappa_{\Df}=\frac{\Gamma(2\Df)^4}{2\Gamma(\Df)^4 \Gamma(4\Df-1)}
\ea
Hence we conclude that the super sum rule must be
\ba
\sum_{\Delta} a_{\Delta} \tilde \beta_0(\Delta)=\frac{g_4}{8\pi \kappa_{\Df}}
\ea
This is true at tree level. Non-perturbatively we can define an effective coupling via:
\ba
\mathcal G(z)-1\underset{z\to \infty}=\frac{i\pi}{z}\, \frac{g_4^{\mbox{\tiny eff}}}{8\pi(\Df-\frac 12)^2}\,, \quad \mbox{or equivalently,}\quad n^2 \gamma_n\underset{n\to \infty}=\frac{g_4^{\mbox{\tiny eff}}}{8\pi}
\ea
in which case the sum rule is
\ba
\sum_{\Delta} a_{\Delta} \tilde \beta_0(\Delta)=\frac{g_4^{\mbox{\tiny eff}}}{8\pi \kappa_{\Df}}
\ea

\section{Asymptotic freedom from Regge bound}
\label{app:freedom}

    Suppose we have a correlator of a field of dimension $\Df$. 
     We assume that 
    \ba
    \ud^2 \mathcal G(z):=(1-z)^{2\Df}\mathcal G(z)-\mathcal R_z\mathcal G(\tfrac{z}{z-1}) \underset{z\to 1}\sim \nu (1-z)^{1-\tj}
    \ea
    Using the OPE we have
    \ba
    \ud^2 \mathcal G(1-z)=\frac{z^{2\Df}}{(1-z)^{2\Df}}\int_0^{\infty}\, \ud\mu_{\Delta} \, 2\, \sin^2[\tfrac{\pi}2 (\Delta-2\Df)]\,  \hat G_\Delta(1-z)\,,\label{d2ope}
    \ea
with
\ba
\ud \mu_\Delta:=\ud \Delta\, \frac{a_\Delta}{a_{\Delta}^{\tt gff}}\,, \qquad \hat G_\Delta(z)=\, a_{\Delta}^{\tt gff}\, z^{\Delta}\, _2F_1(\Delta,\Delta,2\Delta,z)
\ea
In the limit $z\to 0$ individual blocks in \reef{d2ope} behave as $z^{2\Df+2n}[a_n \log(z)+b_n]$. It is convenient to assume that $1-\tj<2\Df$ so that such behaviour is subleading relative to the overall behaviour, but our results do not depend on this assumption.\footnote{The point is that we can always make the $(1-\tj)$ power law behaviour leading by working with the function
\ba
d^2_p\mathcal G(z):= (\tfrac{1-z}{z})^{2\Df}(C^2_z)^p\left[(\tfrac{z}{1-z})^{2\Df}\ud^2 \mathcal G(z)\right] \sim \nu_p (1-z)^{1-\tj-p}\,.
\ea
with $C^2_z$ the $SL(2,R)$ Casimir operator which acts diagonally on blocks, $C^2_z G(z,\Delta)\propto G(z,\Delta)$.
}

We are interested in the limit $|z|\to 0$. Let us therefore introduce some scale $1\ll \Delta^*\ll (1-z)^{-\frac 12}$ where we split the sum over states. We have
\ba
G(1-z,\Delta)\underset{\Delta,(1-z)^{-\frac 12}\gg 1}=z^{\frac 12-2\Df}\, S(\Delta \sqrt{z})\,, \qquad S(x)=\frac{8}{\Gamma(2\Df)^2}\, x^{4\Df-1}\, K_0(2 x)
\ea
up to $O(1/\Delta)$ corrections. Note that
\ba
S(x)\geq 0\,\qquad S(x)\underset{x\ll 1}=O(x^{4\Df-1}\log(x))\,, \qquad S(x)\underset{x\gg 1}=O(e^{-x})
\ea
We can now write
\ba
\ud^2 \mathcal G(1-z)\underset{z \to 0}\sim\sqrt{z}\int_{\Delta^*}^\infty \ud \mu_\Delta\, 2\sin^2[\tfrac{\pi}2(\Delta-2\Df)]\, S(\Delta\sqrt{z}) \label{d2tmp}
\ea
Introducing
\ba
B_n=[\Delta_n-1,\Delta_n+1]\,, \qquad B_n^{\epsilon}=[\Delta_n-\epsilon,\Delta_n+\epsilon]
\ea
we get from \reef{d2tmp}
\ba
\sum_{n=N}^\infty \int_{B_n} \ud \mu_\Delta\, 2\sin^2[\tfrac{\pi}{2}(\Delta-2\Df)] S(\Delta\sqrt{z})\sim \nu z^{\frac 12-\tj}
\ea
But now note that the left-hand side satisfies
\ba
\sum_{n=N}^\infty \int_{B_n} \ud \mu_\Delta\, 2\sin^2[\tfrac{\pi}{2}(\Delta-2\Df)] S(\Delta\sqrt{z})\geq C z^{2\Df-\frac 12}
\sum_{n=N}^\infty \langle \gamma_n^2\rangle \Delta_n^{4\Df-1} e^{-\Delta_n\sqrt{z}}
\ea
with
\ba
\langle \gamma_n^2\rangle:=\frac{4}{\pi^2}\int_{B_n} \ud \mu_\Delta\, \sin^2[\tfrac{\pi}{2}(\Delta-2\Df)] 
\ea
The Hardy-Littlewood Tauberian theorem (see \cite{Pappadopulo:2012jk} for a physics friendly summary) now implies
%
%
%
%
\ba
\langle \gamma_n^2\rangle=O(n^{2\tj-1})
\ea
Note that this is rigorous but relatively weak, following from $\sum^n \langle \gamma_k^2\rangle \approx n^{2\tj-1}$ 
Roughly we actually expect the stronger $\langle \gamma_n^2\rangle \approx n^{2\tj-2}$.
In any case, for $\tj<\frac 12$ anomalous dimensions must necessarily decay down to zero for large $n$. To formalize this, note
we get easily that for any $s,\epsilon<s$
\ba
\int_{\Delta_n+s-\epsilon}^{\Delta_n+s+\epsilon} \ud \mu_{\Delta}=O(n^{2 \tj-1})
\ea
i.e. for any interval not containing a $\Delta_n$ the OPE contributions must vanish. Conversely, from (see \cite{Mazac:2018ycv} for a partial proof):
\ba
\lim_{n\to \infty}\int_{B_n} \ud \mu_{\Delta}=1\quad \Rightarrow \quad \lim_{n\to \infty}\int_{\Delta_n-\epsilon}^{\Delta_n+\epsilon} \ud \mu_{\Delta}=1
\ea
In fact we can promote this to 
\ba
\lim_{n\to \infty}\int_{\Delta_n-\delta_n}^{\Delta_n+\delta_n} \ud \mu_{\Delta}=1
\ea
with $\delta_n\approx n^{\frac 12-\tj-\epsilon}$.
\footnote{This follows from considering the converse region where we lower bound the result by approximating the $\sin^2$ as $\epsilon_n^2$}
These results imply
\ba
 \frac{\int_{\Delta_n-\delta_n}^{\Delta_n+\delta_n} \ud \mu_\Delta (\Delta-\Delta_n)^2}{\int_{\Delta_n-\delta_n}^{\Delta_n+\delta_n} \ud \mu_\Delta}=O(n^{2\tj-1})\,, \quad 
 \frac{\int_{\Delta_n-\delta_n}^{\Delta_n+\delta_n} \ud \mu_\Delta (\Delta-\Delta_n)}{\int_{\Delta_n-\delta_n}^{\Delta_n+\delta_n} \ud \mu_\Delta}=O(n^{\tj-\frac 12})
\ea

\section{OPE coefficients of quadruple-twist operators} \label{Appendix: twist4}

In this appendix, we study the $\epsilon$-expansion of the following OPE in the one-dimensional long-range Ising model:
\begin{equation}\label{eq: quad}
    \phi(x)\phi(0)
    \sim
    x^{\Delta_n - 2\Delta_\phi}\,
    \lambda_{\phi\phi[\phi^4]_n}\,[\phi^4]_n(0),
    \qquad
    \Delta_n := 4\Delta_\phi + 2n.
\end{equation}
The OPE coefficients $\lambda_{\phi\phi[\phi^4]_n}$ first appear at order $\epsilon^2$, while their squares $\lambda_{\phi\phi[\phi^4]_n}^2$ enter the $\epsilon^4$ correction to the four-point crossing equation discussed in the main text.

A key subtlety arises from operator degeneracy.  
For the double-twist operators $[\phi^2]_n$, each level $n$ (corresponding to $2n$ derivatives) contains a unique primary operator.  
By contrast, the quadruple-twist tower $[\phi^4]_n$ is degenerate at fixed $n$.\footnote{
The number of primary operators at level $n$ is
\[
\frac{1}{72}\left(6n(n+6) + 9(-1)^n + 16\cos\!\left(\frac{2\pi n}{3}\right) + 47\right).
\]
For example, starting from $n=0$, the degeneracies at the first few levels are
\[
1,\,1,\,2,\,3,\,4,\,5,\,7,\,8,\,\ldots.
\]
}

In the GFF theory there is no canonical way to resolve this degeneracy.  
However, in the interacting long-range Ising model, the $\phi^4$ interaction lifts the degeneracy: operators $[\phi^4]_{n,i}$ at fixed level $n$ acquire distinct anomalous dimensions $\gamma_{n,i}$.  
In practice, one diagonalizes the two-point and three-point functions simultaneously, imposing\footnote{See the discussion around Eq.~\eqref{eq: quad-anorm} about the $\phi^4$ OPE and anomalous dimension.}
\[
\langle [\phi^4]_{n,i} [\phi^4]_{n,j} \rangle \sim \delta_{ij},
\qquad
\langle [\phi^4]_{n,i} [\phi^4]_{n,j} \phi^4 \rangle \sim \delta_{ij},
\]
so that the operators are orthonormal and have definite anomalous dimensions.

The individual OPE coefficients $\lambda_{\phi\phi[\phi^4]_{n,i}}$ depend on the choice of basis within the degenerate subspace.  
The physically meaningful, basis-independent quantity is therefore the summed OPE coefficient
\[
\lambda_{\phi\phi[\phi^4]_n}^2
=
\sum_i
\lambda_{\phi\phi[\phi^4]_{n,i}}^2.
\]
Our task in this appendix is to compute this summed quantity.

A direct calculation of such quantities from Feynman-diagram is complicated, since it involves multi-loop calculation and a delicate resolution of the degeneracy at different level. Thus we employ a detour strategy to evaluate this value. We begin by noticing the following equation from the non-local equation of motion in long-range Ising\cite{Paulos:2015jfa}:
\begin{equation}
    \frac{\lambda_{\phi \mathcal O_1 \mathcal O_2}}{\lambda_{\phi^3 \mathcal O_1 \mathcal O_2}}=\frac{\lambda_{\phi \mathcal O_3 \mathcal O_4}}{\lambda_{\phi^3 \mathcal O_3 \mathcal O_4}}\, \times \frac{\Gamma \left(\frac{-\Delta_{34}-\Df +1}{2} \right) \Gamma \left(\frac{\Delta_{34}-\Df
   +1}{2} \right) \Gamma \left(\frac{\Df -\Delta_{12}}{2}\right) \Gamma \left(\frac{\Delta_{12}+\Df
   }{2}\right)}{\Gamma \left(\frac{\Df -\Delta_{34}}{2}\right) \Gamma \left(\frac{\Delta_{34}+\Df
   }{2}\right) \Gamma \left(\frac{-\Delta_{12}-\Df +1}{2} \right) \Gamma \left(\frac{\Delta_{12}-\Df +1}{2}
   \right)}
\end{equation}
Here $\Delta_{ij}=\Delta_i-\Delta_j$. We set
\[
\mathcal O_1=\mathcal O_3=\phi,
\qquad
\mathcal O_2=[\phi^4]_{n,i},
\qquad
\mathcal O_4=\phi^2,
\]
and use the $\epsilon$-expansion data
\begin{align}
\lambda_{\phi\phi\phi^2}
    &= \sqrt{2}+O(\epsilon),
&
\lambda_{\phi^3\phi\phi^2}
    &= \sqrt{3}+O(\epsilon),\\[4pt]
\Delta_{\phi^2}
    &= 2\Delta_\phi + \frac{\epsilon}{3}
       + O(\epsilon^2),
&
\Delta_{[\phi^4]_n}
    &= 4\Delta_\phi + 2n + \gamma_{n,i}\epsilon+ O(\epsilon^2).
\end{align}

This yields
\begin{equation}
    \lambda_{\phi\phi[\phi^4]_{n,i}}^2
    =
    \epsilon^4
    \frac{\pi^3}{54}
    \frac{\Gamma(2n+1)^2}
         {\Gamma\!\left(2n+\tfrac{3}{2}\right)^2}
    (\gamma_{n,i}-1)^2\,
    \lambda_{\phi^3\phi[\phi^4]_{n,i}}^2.
\end{equation}

To determine the anomalous dimensions $\gamma_{n,i}$, we employ the standard OPE method.  
Consider the operator product
\[
[\phi^4]_{n,i}(x)\times \phi^4(0)
\sim
\frac{\lambda_{[\phi^4]_{n,i},[\phi^4]_{n,i}\phi^4}}
{|x|^{1-\epsilon}}\,
[\phi^4]_{n,i}(0).
\]
Integrating this expression over spacetime produces a $1/\epsilon$ pole, which must be cancelled by wavefunction renormalization.  More precisely, the renormalization constant takes the form
\begin{equation}
    Z_{n,i}
    =
    1
    -
    \frac{
    \lambda_{[\phi^4]_{n,i},[\phi^4]_{n,i}\phi^4}
    \,\lambda^* S_d
    }{
    4!\,\epsilon
    }.
\end{equation}

Here $\lambda^*$ is the fixed-point coupling~\cite{Paulos:2015jfa},
\begin{equation}
\lambda^*=\frac{2}{3S_d}\,\epsilon,
\end{equation}
and $S_d$ denotes the volume of the unit sphere in $d$ dimensions (in $d=1$, $S_1=2$).

The anomalous dimension is then extracted from the logarithmic divergence of the renormalization constant, yielding
\begin{equation}\label{eq: quad-anorm}
\gamma_{n,i}
=
\frac{1}{36}\,
\lambda_{[\phi^4]_{n,i},[\phi^4]_{n,i}\phi^4}.
\end{equation}

Combine everything, we have:
\begin{equation}
    \lambda_{\phi \phi [\phi^4]_{n}}^2=\sum_i\lambda_{\phi \phi [\phi^4]_{n,i}}^2= \epsilon^4\frac{\pi ^3}{54}\frac{ \Gamma (2 n+1)^2}{\Gamma \left(2
   n+\frac{3}{2}\right)^2}\sum_i\,
   \left(\frac{ 1}{36}\lambda_{[\phi^4]_{n,i},[\phi^4]_{n,i}\phi^4 }-1\right)^2\lambda_{\phi^3 \phi [\phi^ 4]_{n,i}}^2 
\end{equation}

All quantities on the right-hand side are tree-level OPE data and can be computed by Wick contractions. Practical computations up to $n\simeq 50$ reveal a remarkably simple closed-form result:
\begin{equation}
\lambda_{\phi\phi[\phi^4]_n}^2
=
\frac{
\pi^2 2^{1-8n}
\left(
2\,\Gamma\!\left(n+\tfrac{1}{2}\right)^2
+
\pi\,\Gamma(n+1)^2
\right)
\Gamma(4n+1)
}{
81\,\Gamma(n+1)^2
\Gamma\!\left(2n+\tfrac{3}{2}\right)^2
}
\epsilon^4.
\end{equation}

Finally, normalizing by the generalized free field coefficient $a^{\text{gff}}_{1+2n}$,
\begin{equation}
\frac{\lambda_{\phi\phi[\phi^4]_n}^2}
     {a^{\text{gff}}_{1+2n}}
=
\epsilon^4
\left(
\frac{\pi^3}{162\,n}
-
\frac{(\pi-8)\pi^2}{648\,n^2}
+
\frac{(\pi-16)\pi^2}{2592\,n^3}
-
\frac{(\pi-20)\pi^2}{10368\,n^4}
+ \cdots
\right).
\end{equation}

\section{Epsilon expansion for long range Lee-Yang model}
\label{app: Lee-Yang}

In this section we will show the results of the epsilon expansion of the Lee-Yang model up to the $\epsilon^2$ order. The model is defined by the Wilson-Fisher fixed point of the following bare Lagrangian:
\begin{equation}
\mathcal{L}= \psi \partial^s \psi+\frac{1}{3!} g_0 \psi^3
\end{equation}

We remind the reader that the epsilon expansion is defined as:
\begin{equation}
    \Delta_\phi(\epsilon)=\frac{1}{3}-\frac{1}{3}\epsilon
\end{equation}
which corresponds to $s=\frac{1}{3}+\frac{2}{3}\epsilon$. To begin with, we need the renormalized Lagrangian:
\begin{equation}
    \mathcal{L}= \psi \partial^s \psi+\frac{1}{3!} g Z_g \psi^3
\end{equation}
where $Z_g=1+\delta_g$ is the renormalized factor of the coupling constant, with the epsilon expansion of the $\delta_g$ given by:
\begin{equation}
    \delta_g=\frac{f_1(g)}{\epsilon}+\frac{f_2(g)}{\epsilon^2}+\cdots
\end{equation}

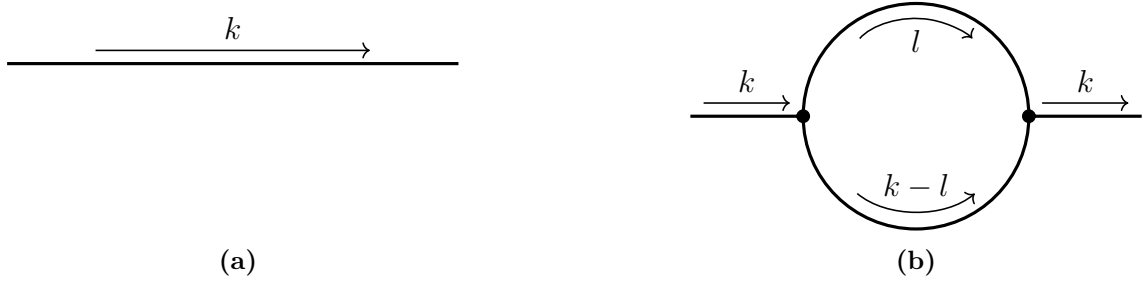
\begin{figure}[t!]
    \centering
    \begin{subfigure}{0.45\textwidth}
      \centering
      \begin{tikzpicture}[very thick, q0/.style={->,semithick,yshift=5pt,shorten >=5pt,shorten <=5pt}]
        \draw (-3,-2) -- (3,-2);
        
        \draw[q0] (-2,-2) -- (2,-2) node[midway,above] {$k$};
    \end{tikzpicture}
    \vspace{2.2cm}
      \caption{}\label{fig: two-pointtree1}
    \end{subfigure}
    \hfill
    \begin{subfigure}{0.45\textwidth}
      \centering
      \begin{tikzpicture}[very thick, q0/.style={->,semithick,yshift=5pt,shorten >=5pt,shorten <=5pt}]
        \def\radius{1.5}
        \draw (0,0) circle (\radius);
        
        \draw[q0] (140:0.75*\radius) arc (140:40:0.75*\radius) node[midway,below] {$l$};
        \draw[q0] (-130:0.95*\radius) arc (-130:-50:0.95*\radius) node[midway,above] {$k-l$};
        \filldraw (-2*\radius,0) -- (-\radius,0) circle (2pt)
                  (\radius,0) circle (2pt) -- (2*\radius,0);
        \draw[q0] (-2*\radius,0) -- (-\radius,0) node[midway,above] {$k$};
        \draw[q0] (\radius,0) -- (2*\radius,0) node[midway,above] {$k$};
      \end{tikzpicture}
      \caption{}\label{fig: two-pointloop1}
    \end{subfigure}
    \caption{  Tree-level and one-loop diagrams for the two-point correlator.}\label{fig: two-point}
  \end{figure}

The standard textbook result for the $\beta$ function reads\footnote{The reader is referred to e.g. \cite{Srednicki:2007qs} for a derivation}:
\begin{equation}\label{eq: beta}
    \beta(g)=-\epsilon g +g^2 f_1'(g)
\end{equation}
We further notice that we don't need a counterterm for the wave function renormalization, since the bare field has a non-local kinetic term. The loop contribution to the two-point correlator must be finite. This can be verified by directly calculating the loop contribution to the two-point correlator at one loop order. In Figure~\ref{fig: two-point} we show the tree-level~\ref{fig: two-pointtree1} and one-loop~\ref{fig: two-pointloop1} diagrams for the two-point correlator. The tree-level diagram is given by\footnote{Here and henceforth in this appendix, we slightly abuse the notation by using $k^s$ to denote $|k|^{s}$ or $(k^2)^{\frac{s}{2}}$, for the sake of brevity.}:
\begin{equation}
    \frac{1}{k^s}
\end{equation}
with the corresponding position space free correlator as its Fourier transform:
\begin{equation}
  \frac{\alpha_s}{x^{2 \Delta_\phi(\epsilon)}}, \quad \alpha_s=\frac{\Gamma(1-s) \sin\left(\frac{\pi s}{2}\right)}{\pi}
\end{equation} 
The one-loop diagram is given by:
\begin{equation}
        \frac{1}{2}(-g)^2 \int \frac{d l}{2 \pi} \frac{1}{l^s} \frac{1}{(k-l)^s} \frac{1}{k^{2 s}} =\frac{1}{2} g^2 I_{c2}\left(\frac{s}{2}, \frac{s}{2}\right)\left(k\right)^{1-4 s}
\end{equation}
which is finite for $s<\frac{1}{2}$. Up to the $g^2$ order, we can thus write:
\begin{equation}
    \langle \tilde{\psi}(k) \tilde{\psi}(-k) \rangle = \frac{1}{k^s}\left(1-\frac{ \pi g^2 }{\Gamma \left(\frac{1}{3}\right)^3}\right)
\end{equation}
By doing a inverse Fourier transform, we can find the two-point correlator in position space:
\begin{equation}
    \langle \psi(x) \psi(0) \rangle = \frac{\mathcal{N}_s}{x^{2 \Delta_\phi}}, \quad \mathcal{N}_s=\left(1-\frac{ \pi g^2 }{\Gamma \left(\frac{1}{3}\right)^3}\right)\frac{\Gamma(1-s) \sin\left(\frac{\pi s}{2}\right)}{\pi}
\end{equation}
So we have:
\begin{equation}\label{eq: normalizedfield}
    \phi(x)=\frac{1}{\sqrt{\mathcal{N}_s}}\psi(x)
\end{equation}
as the normalized primary field studied in the main text\footnote{The dependence of coupling $g$ on $\epsilon$ will be fixed by the vanishing of the beta function at two loop order.}.

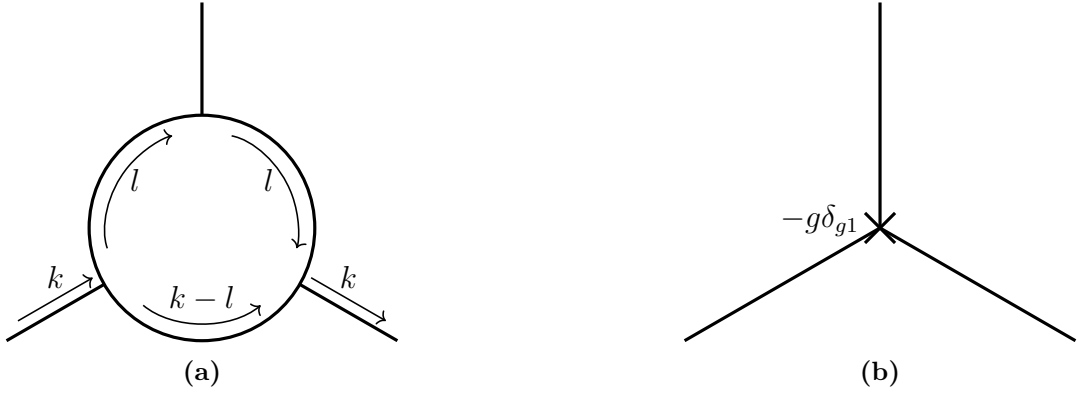
\begin{figure}[t!]
    \centering
    \begin{subfigure}{0.45\textwidth}
      \centering
      \begin{tikzpicture}[very thick, q0/.style={->,semithick,yshift=5pt,shorten >=5pt,shorten <=5pt}, q1/.style={->,semithick,shorten >=5pt,shorten <=5pt}]
        \def\radius{1.5}
        \draw (0,0) circle (\radius);
        
        \draw[q1] (-160:0.85*\radius) arc (-160:-260:0.85*\radius) node[midway,right] {$l$};
        \draw[q1] (80:0.85*\radius) arc (80:-20:0.85*\radius) node[midway,left] {$l$};
        \draw[q0] (-130:0.95*\radius) arc (-130:-50:0.95*\radius) node[midway,above] {$k-l$};
        \filldraw (-150:2*\radius) -- (-150:1.*\radius) ;
        \filldraw (-30:2*\radius) -- (-30:1.*\radius) ;
        \filldraw (90:2*\radius) -- (90:1.*\radius) ;
        \draw[q0] (-150:2*\radius) -- (-150:1*\radius) node[midway,above] {$k$};
        \draw[q0] (-30:1*\radius) -- (-30:2*\radius) node[midway,above] {$k$};
        \end{tikzpicture}
      \caption{}\label{fig: three-pointloop1}
    \end{subfigure}
    \hfill
    \begin{subfigure}{0.45\textwidth}
      \centering
      \begin{tikzpicture}[very thick, 
        q0/.style={->,semithick,yshift=5pt,shorten >=5pt,shorten <=5pt}, 
        q1/.style={->,semithick,shorten >=5pt,shorten <=5pt}]
        
        \def\radius{1.5}
        
        \filldraw (-150:2*\radius) -- (-150:0) ;
        \filldraw (-30:2*\radius) -- (-30:0) ;
        \filldraw (90:2*\radius) -- (90:0) ;
        
        \draw (-0.2,0.2 ) -- (0.2,-0.2);
        \draw (-0.2,-0.2) -- (0.2,0.2);
        
        \node at (-.8,0.1) {$-g\delta_{g1}$};
    
    \end{tikzpicture}
      \caption{}\label{fig: three-pointcounter1}
    \end{subfigure}
    \caption{ One-loop diagrams for the three-point vertex function, whose divergent part will be canceled by the counterterm.}\label{fig: three-pointoneloop}
  \end{figure}

  Now we turn to the calculation of the beta function. From \eqref{eq: beta}, we only need to calculate the divergent part of the amputated three-point vertex function. In Figure~\ref{fig: three-pointoneloop} we show the one-loop diagrams for the three-point vertex function~\ref{fig: three-pointloop1}, whose divergent part is absorbed by the counterterm~\ref{fig: three-pointcounter1}. Since the counterterm here independent of the external momenta, we can directly read off the divergent part of the amputated three-point vertex function from the Figure~\ref{fig: three-pointoneloop} by setting the one of the external momenta to zero:
  \begin{equation}
    \begin{aligned}
        &(-g)^3 \int \frac{d l}{2 \pi} \frac{1}{l^{2 s}} \frac{1}{(k-l)^s} \\
        & =(-g)^3 I_{c_2}\left(s, \frac{s}{2}\right)\left(k^2\right)^{\frac{1}{2}-\frac{3}{2}\epsilon } \sim \frac{(-g)^3}{2 \pi  \epsilon }+O\left(\epsilon ^0\right) 
        \end{aligned}
  \end{equation}
  Therefore the leading order counterterm is given by\footnote{Here and henceforth we will be using the Minimal Subtraction scheme.}:
  \begin{equation}
    \delta_{g_1}=-g^2 \cdot \frac{1}{2 \pi \epsilon} 
  \end{equation}

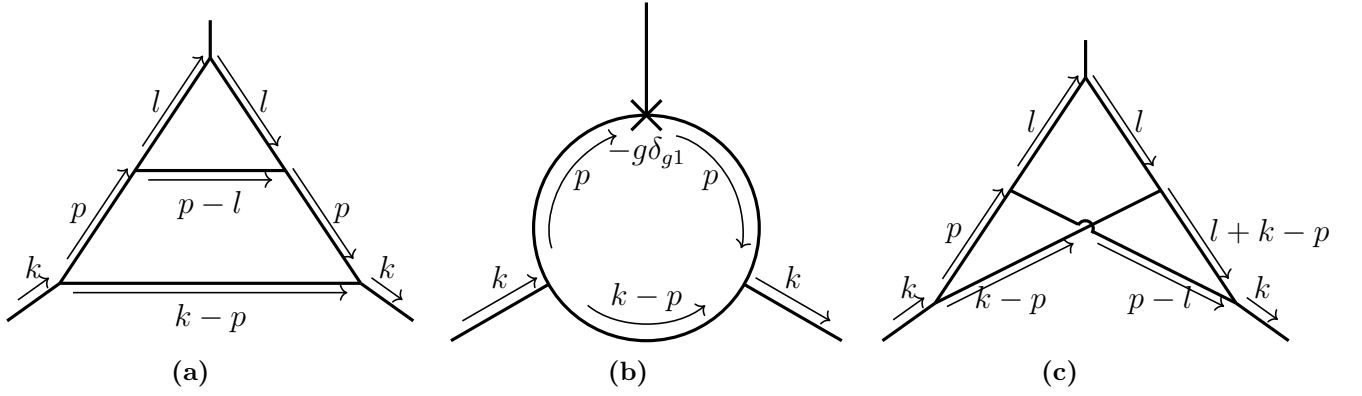
\begin{figure}[t!]
    \centering
    \begin{subfigure}{0.3\textwidth}
      \centering
      \begin{tikzpicture}[very thick, q0/.style={->,semithick,yshift=5pt,shorten >=5pt,shorten <=5pt}, q1/.style={->,semithick,shorten >=5pt,shorten <=5pt}]
        \def\radius{1.5}
        \filldraw (-2,-1) -- (2,-1) ;
        \filldraw (-2,-1) -- (0,2) ;
        \filldraw (0,2) -- (2,-1) ;
        \filldraw (-1,.5) -- (1,.5) ; 
        \filldraw (0,2) -- (0,2.5) ;
        \filldraw (-2,-1) -- (-2.7,-1.5) ; 
        \filldraw (2,-1) -- (2.7,-1.5) ; 
        \draw[q0] (-2.7,-1.5) -- (-2,-1) node[midway,above] {$k$};
        \draw[q0] (2,-1) -- (2.7,-1.5) node[midway,above] {$k$};
        \draw[q0] (-1,.5) -- (0,2) node[midway,left] {$l$};
        \draw[q0] (0,2) -- (1,.5) node[midway,right] {$l$};
        \draw[q0] (-2,-1) --(-1,.5) node[midway,left] {$p$};
        \draw[q0] (1,.5)-- (2,-1) node[midway,right] {$p$};
        \draw[q0] (-2,-1-.3)-- (2,-1-.3) node[midway,below] {$k-p$};
        \draw[q0] (-1,.5-.3) -- (1,.5-.3) node[midway,below] {$p-l$};
        \end{tikzpicture}
      \caption{}\label{fig: three-pointloop21}
    \end{subfigure}
    \hfill
    \begin{subfigure}{0.29\textwidth}
      \centering
      \begin{tikzpicture}[very thick, q0/.style={->,semithick,yshift=5pt,shorten >=5pt,shorten <=5pt}, q1/.style={->,semithick,shorten >=5pt,shorten <=5pt}]
        \def\radius{1.5}
    
        \draw (0,0) circle (\radius);
        
        \draw[q1] (-160:0.85*\radius) arc (-160:-260:0.85*\radius) node[midway,right] {$p$};
        \draw[q1] (80:0.85*\radius) arc (80:-20:0.85*\radius) node[midway,left] {$p$};
        \draw[q0] (-130:0.95*\radius) arc (-130:-50:0.95*\radius) node[midway,above] {$k-p$};
        \filldraw (-150:2*\radius) -- (-150:1.*\radius) ;
        \filldraw (-30:2*\radius) -- (-30:1.*\radius) ;
        \filldraw (90:2*\radius) -- (90:1.*\radius) ;
        \draw[q0] (-150:2*\radius) -- (-150:1*\radius) node[midway,above] {$k$};
        \draw[q0] (-30:1*\radius) -- (-30:2*\radius) node[midway,above] {$k$};
        \draw (-0.2,0.2+\radius) -- (0.2,-0.2+\radius);
        \draw (-0.2,-0.2+\radius) -- (0.2,0.2+\radius);
        \node at (0,-0.5+\radius) {$-g\delta_{g1}$};
        \end{tikzpicture}
      \caption{}\label{fig: three-pointcounter2}
    \end{subfigure}
    \hfill
    \begin{subfigure}{0.29\textwidth}
        \centering
        \begin{tikzpicture}[very thick, q0/.style={->,semithick,yshift=5pt,shorten >=5pt,shorten <=5pt}, q1/.style={->,semithick,shorten >=5pt,shorten <=5pt}]
            \def\radius{1.5}
            
            \filldraw (-2,-1) -- (0,2) ;
            \filldraw (0,2) -- (2,-1) ;
            \filldraw (-2,-1) -- (1,.5) ;
            \filldraw (-1,.5) -- (-.1,.05) ;
            \filldraw (.1,-.05) -- (2,-1)  ; 
            \draw (160:0.07*\radius) arc (155:-35:0.07*\radius) ; 
            \filldraw (0,2) -- (0,2.5) ;
            \filldraw (-2,-1) -- (-2.7,-1.5) ; 
            \filldraw (2,-1) -- (2.7,-1.5) ; 
            \draw[q0] (-2.7,-1.5) -- (-2,-1) node[midway,above] {$k$};
            \draw[q0] (2,-1) -- (2.7,-1.5) node[midway,above] {$k$};
            \draw[q0] (-1,.5) -- (0,2) node[midway,left] {$l$};
            \draw[q0] (0,2) -- (1,.5) node[midway,right] {$l$};
            \draw[q0] (-2,-1) --(-1,.5) node[midway,left] {$p$};
            \draw[q0] (1,.5)-- (2,-1) node[midway,right] {$l+k-p$};

            \draw[q0] (-2,-1-.3)-- (0,0-.3) node[midway,below] {$k-p$};
            \draw[q0] (0,0-.3)--(2,-1-.3)  node[midway,below] {$p-l$};
        \end{tikzpicture}
        \caption{}\label{fig: three-pointloop22}
      \end{subfigure}
    \caption{  Tree-level and one-loop diagrams for the position space three-point correlators.}\label{fig: three-pointtwoloop}
\end{figure}

  The convergent part of the two-loop diagrams is given by Figure~\ref{fig: three-pointtwoloop}. Multiplying the symmetry factor, the diagram \ref{fig: three-pointloop21} and \ref{fig: three-pointcounter2} are given by:
\begin{equation}\label{eq: three-pointtwoloop1}
    \begin{aligned}
        & 3(-g)^2 \int \frac{d p}{2 \pi} \frac{1}{p^{2 s}} \frac{1}{(k-p)^s}\left(-g \delta_{g_1}+(-g)^3 \int \frac{d l}{2 \pi} \frac{1}{l^{2 s}} \frac{1}{(p-l)^s}\right) \\
        &\sim 3(-g)^5\left(-\frac{1}{8 \pi ^2 \epsilon ^2}+\frac{2 \sqrt{3} \pi +9 \log (3)}{24 \pi ^2 \epsilon }\right)
      \end{aligned}
\end{equation}
The most non-trivial part is the diagram \ref{fig: three-pointloop22}, which is given by:
\begin{equation}
  \begin{aligned}
    & =\frac{1}{2}(-g)^5 \int \frac{d l}{2 \pi} \frac{d p}{2 \pi} \frac{1}{l^{2 s}} \frac{1}{(k-p)^s} \frac{1}{(p-l)^s} \frac{1}{\left(l+k-p\right)^s} \frac{1}{p^s } \\
    & =\frac{1}{2}(-g)^5 \int \frac{d p}{2 \pi} \frac{1}{p^s} \frac{1}{\left(k-p\right)^s} I_3\left(s, \frac{s}{2}, \frac{s}{2},-p, k-p\right) 
    \end{aligned}
\end{equation}
Here we notice that there is a symmetry factor of $\frac{1}{2}$ in the diagram \ref{fig: three-pointloop22}, which is due to the fact that the diagram is symmetric under the exchange of internal vertices. The $I_3$ function defined in Appendix~\ref{App: Feynman integrals} and the factor $I_3\left(s, \frac{s}{2}, \frac{s}{2},-p, k-p\right) $ is convergent in the neighborhood of $s=\frac{1}{3} $. Therefore, the divergent piece of the diagram \ref{fig: three-pointloop22} is given by:
\begin{equation}\label{eq: three-pointtwoloop2}
  \begin{aligned}
    & \sim \frac{1}{2}(-g)^5 \int \frac{d p}{2 \pi} \frac{1}{p^s} \frac{1}{(p-k)^s}\left(p^2\right)^{\frac{1}{2}-2 s} I_{c_2}(s, s) \\
    & =\frac{1}{2}(-g)^5\left(k^2\right)^{1-3 s} I_{c_2}\left(\frac{5}{2} s-\frac{1}{2}, \frac{s}{2}\right)  I_{c_2}(s, s) . \\
    & \sim (-g)^5\frac{3 \sqrt{3} \Gamma^3\left(\frac{1}{3}\right)}{32 \pi^3} \frac{1}{\epsilon}
    \end{aligned}
\end{equation}

\begin{figure}[t!]
    \centering
    \begin{tikzpicture}[very thick, q0/.style={->,semithick,yshift=5pt,shorten >=5pt,shorten <=5pt}, q1/.style={->,semithick,shorten >=5pt,shorten <=5pt}]
      \def\radius{1.5}
      \filldraw (-2,-1) -- (-.6,-1) ;
      \filldraw (.6,-1) -- (2,-1)  ;
      \filldraw (-2,-1) -- (0,2) ;
      \filldraw (0,2) -- (2,-1) ;
      \def\radius{1.5}
      \draw (0,-1) circle (.4*\radius); 
      \filldraw (0,2) -- (0,2.5) ;
      \filldraw (-2,-1) -- (-2.7,-1.5) ; 
      \filldraw (2,-1) -- (2.7,-1.5) ; 
      \draw[q0] (-2.7,-1.5) -- (-2,-1) node[midway,above] {$k$};
      \draw[q0] (2,-1) -- (2.7,-1.5) node[midway,above] {$k$};
      \draw[q0] (-2,-1) -- (0,2) node[midway,left] {$l$};
      \draw[q0] (0,2) -- (2,-1) node[midway,right] {$l$};

      \end{tikzpicture}
    \caption{}\label{fig: three-pointloop3}
\end{figure}

The last divergent piece of the two-loop diagrams are given by Figure~\ref{fig: three-pointloop3}, which is given by:
\begin{equation}\label{eq: three-pointtwoloop3}
    3(-g)^5 \int \frac{dl}{2 \pi} \frac{1}{l^{2s}} \frac{1}{2} I_{c_2}\left(\frac{s}{2}, \frac{s}{2}\right)\left((k-l)^2\right)^{\frac{1}{2}-2s} \sim-\frac{3}{4}(-g)^5 \frac{1}{\Gamma\left(\frac{1}{3}\right)^3 \epsilon}
\end{equation}
Here we reused the result of previous two-point correlator calculation for Figure~\ref{fig: two-point}.

Combining all the divergent pieces \eqref{eq: three-pointtwoloop1}, \eqref{eq: three-pointtwoloop2} and \eqref{eq: three-pointtwoloop3}, the divergent part of the two-loop diagrams can be absorbed by the following counterterm:
\begin{equation}
  \delta_{g_2}=\frac{3 g^4}{8 \pi ^2 \epsilon ^2}-\frac{g^4 \left(-24 \pi ^3+8 \sqrt{3} \pi ^2 \Gamma \left(\frac{1}{3}\right)^3+3 \sqrt{3} \Gamma \left(\frac{1}{3}\right)^6+36 \pi  \log (3) \Gamma \left(\frac{1}{3}\right)^3\right)}{32 \pi ^3   \Gamma \left(\frac{1}{3}\right)^3}\frac{1}{\epsilon}
\end{equation}
From the vanishing of the beta function \eqref{eq: beta}, we can determine the value of the critical coupling constant $g$ as:
\begin{equation}\label{eq: criticalcoupling}
  0=\beta(g)=-g \epsilon-\frac{g^3}{\pi }+g^5 \left(-\frac{\sqrt{3}}{\pi }-\frac{9 \log (3)}{2 \pi ^2}+\frac{3}{\Gamma \left(\frac{1}{3}\right)^3}-\frac{3 \sqrt{3} \Gamma \left(\frac{1}{3}\right)^3}{8 \pi ^3}\right)
\end{equation}

\begin{figure}[t!]
    \centering
    \begin{subfigure}{0.3\textwidth}
      \centering
      \begin{tikzpicture}[very thick, q0/.style={->,semithick,yshift=5pt,shorten >=5pt,shorten <=5pt}, q1/.style={->,semithick,shorten >=5pt,shorten <=5pt},scale=.8]
        \def\radius{1.5}
        \filldraw (-150:2*\radius) -- (-150:0) ;
        \filldraw (-30:2*\radius) -- (-30:0) ;
        \filldraw (90:2*\radius) -- (90:0) ;
        \begin{feynman}
            \vertex [dot] (L1) at (-150:2*\radius) {};
            \vertex [dot] (R1) at (-30:2*\radius) {};
            \vertex [dot] (T1) at (90:2*\radius) {};
            \vertex [dot] (M) at (-150:0) {};
           
            \node at (T1) [above] {0};
            \node at (L1) [left] {1};
            \node at (R1) [right] {$\infty$};

            \node at (M) [right] {$y_1$};
        \end{feynman}
    \end{tikzpicture}
      \caption{}\label{fig: OPE-pointtree1}
    \end{subfigure}
    \hfill
    \begin{subfigure}{0.3\textwidth}
      \centering
      \begin{tikzpicture}[very thick, q0/.style={->,semithick,yshift=5pt,shorten >=5pt,shorten <=5pt}, q1/.style={->,semithick,shorten >=5pt,shorten <=5pt},scale=.8]
        \def\radius{1.5}
        \draw (0,0) circle (\radius);
        \filldraw (-150:2*\radius) -- (-150:1.*\radius) ;
        \filldraw (-30:2*\radius) -- (-30:1.*\radius) ;
        \filldraw (90:2*\radius) -- (90:1.*\radius) ;
        \begin{feynman}
            \vertex [dot] (L1) at (-150:2*\radius) {};
            \vertex [dot] (R1) at (-30:2*\radius) {};
            \vertex [dot] (T1) at (90:2*\radius) {};
            \vertex [dot] (L2) at (-150:1*\radius) {};
            \vertex [dot] (R2) at (-30:1*\radius) {};
            \vertex [dot] (T2) at (90:1*\radius) {};
            \node at (T1) [above] {0};
            \node at (L1) [left] {1};
            \node at (R1) [right] {$\infty$};

            \node at (T2) [below] {$y_1$};
            \node at (L2) [right] {$y_2$};
            \node at (R2) [left] {$y_3$};
        \end{feynman}
    \end{tikzpicture}
      \caption{}\label{fig: OPE-pointloop1}
    \end{subfigure}
    \hfill
    \begin{subfigure}{0.3\textwidth}
      \centering
      \begin{tikzpicture}[very thick, q0/.style={->,semithick,yshift=5pt,shorten >=5pt,shorten <=5pt}, q1/.style={->,semithick,shorten >=5pt,shorten <=5pt},scale=.8]
        \def\radius{1.5}
        \filldraw (-150:2*\radius) -- (-150:0) ;
        \filldraw (-30:2*\radius) -- (-30:0) ;
        \filldraw (90:2*\radius) -- (90:0) ;
        \begin{feynman}
            \vertex [dot] (L1) at (-150:2*\radius) {};
            \vertex [dot] (R1) at (-30:2*\radius) {};
            \vertex [dot] (T1) at (90:2*\radius) {};
            \draw (-0.2,0.2) -- (0.2,-0.2 );
            \draw (-0.2,-0.2) -- (0.2,0.2);
           
            \node at (T1) [above] {0};
            \node at (L1) [left] {1};
            \node at (R1) [right] {$\infty$};
            \node at (M) [below] {$-g\delta_{g1}$};
            
        \end{feynman}
    \end{tikzpicture}
      \caption{}\label{fig: OPE-pointtree3}
    \end{subfigure}
    \caption{  Tree-level and one-loop diagrams for the position space three-point correlators.}\label{fig: OPE-point}
\end{figure}
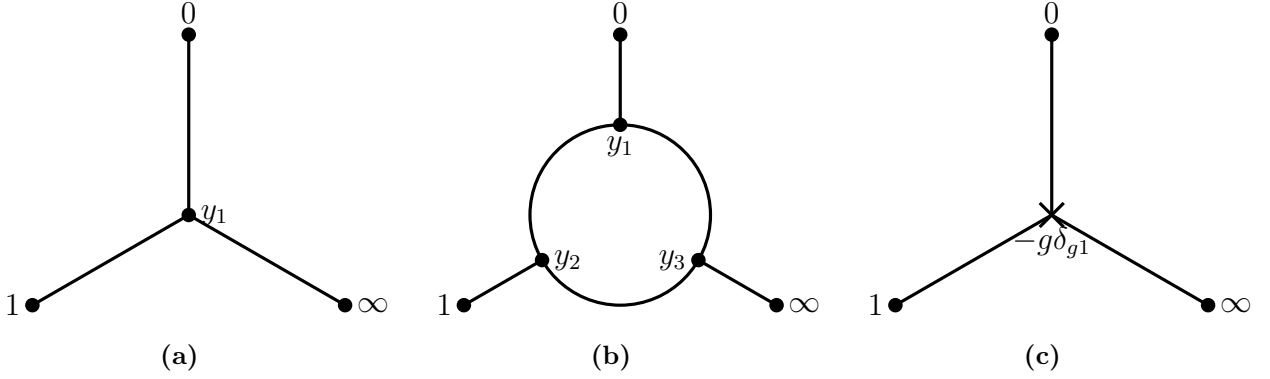

At last we turn to the calculation of the OPE coefficient. What remains is to compute the tree-level and one-loop diagrams for the position space three-point correlators $\langle \psi(0) \psi(1)\psi(\infty) \rangle$, up to a normalization factor defined in Equation~\ref{eq: normalizedfield}. More precisely, we have:
\begin{equation}
    a^2=\langle \phi(0) \phi(1)\phi(\infty) \rangle^2=\frac{1}{\mathcal{N}_s^3} \langle \psi(0) \psi(1)\psi(\infty) \rangle^2
\end{equation}

The tree-level and one-loop diagrams are given by Figure~\ref{fig: OPE-point}. We start by considering the tree-level diagram \ref{fig: OPE-pointtree1}. The corresponding Feynman diagram is given by:
\begin{equation}
  \begin{aligned}
    (-g) \mathcal{N}_s^3 \int d y_1\, y_1^{-2 \Delta_\phi}\left(y_1-1\right)^{-2 \Delta_\phi}=(-2 \pi g) \mathcal{N}_s^3 I_{c_2}(\Delta_\phi, \Delta_\phi) .
    \end{aligned}
\end{equation}
Here we used the normalization factor $\mathcal{N}_s$ defined in Equation~\ref{eq: normalizedfield}, which automatically absorbs the one loop correction to the two-point correlator of the external legs. By the same way, the one-loop diagram \ref{fig: OPE-pointloop1} is given by:
\begin{equation}
  \begin{aligned}
    & (-g)^3 \alpha_s^6 \int d y_1 d y_2 d y_3 y_1^{-2 \Delta_\phi}\left(y_2-1\right)^{-2 \Delta_\phi}\left(y_1-y_2\right)^{-2 \Delta_\phi} \left(y_2-y_3\right)^{-2 \Delta_\phi}\left(y_3-y_1\right)^{-2 \Delta_\phi}\\
    & =(-2\pi g)^3 \alpha_s^6 I_{c_2}\left(\Delta_\phi, \Delta_\phi\right) I_{c_2}(3 \Delta_\phi-\frac{1}{2}, \Delta_\phi) I_{c_2}(4 \Delta_\phi-1, \Delta_\phi).
  \end{aligned}
\end{equation}
and the counterterm is given by:
\begin{equation}
  \begin{aligned}
    \left(-g \delta_{g_1}\right) \alpha_s^3 \int d{y_1} y_1^{-2 \Delta_\phi}\left(y_1-1\right)^{-2 \Delta_\phi} =(-2 \pi g \delta_{g_1}) \alpha_s^3 I_{c_2}(\Delta_\phi, \Delta_\phi) .
  \end{aligned}
\end{equation}
Summing up the tree-level and one-loop diagrams, we get:
\begin{equation}
  \begin{aligned}
    a^2=(2 \pi g)^2 \mathcal{N}_s^{-3} I_{c_2}(\Delta_\phi, \Delta_\phi)^2 \left(\mathcal{N}_s^3+\delta_{g_1}\alpha_s^3+(2 \pi g)^2 \alpha_s^6I_{c_2}(3 \Delta_\phi-\frac{1}{2}, \Delta_\phi) I_{c_2}(4 \Delta_\phi-1, \Delta_\phi)\right)^2.
  \end{aligned}
\end{equation}
Substituting the critical coupling constant $g$ defined in \eqref{eq: criticalcoupling} into the above equation, we get:
\begin{equation}
  \begin{aligned}
    a^2=-\epsilon\frac{3 \sqrt{3}   \Gamma \left(\frac{1}{3}\right)^3 }{4 \pi }\left(1+\frac{  \left(8 \pi ^2+4 \sqrt{3} \pi  \log (27)+9 \Gamma \left(\frac{1}{3}\right)^3\right)}{8 \sqrt{3} \pi }\epsilon\right).
  \end{aligned}
\end{equation}

\section{Some Feynman integrals in long range models}\label{App: Feynman integrals}

In this section we introduce some short hand notation for the Feynman integrals appearing in the epsilon expansion of the long range models. We start by considering the following integral\footnote{In this section, unless otherwise specified, all the integrals are Lebesgue integration on the real line.}:
\begin{equation}
    I_2(\alpha, \beta, k)=\int \frac{d l}{2 \pi} \frac{1}{\left(l^2\right)^\alpha} \frac{1}{\left((l+k)^2\right)^\beta}
\end{equation}
It will be a standard task to compute the closed form of this integral:
\begin{equation}
    I_2(\alpha, \beta, k)=I_{c2}(\alpha, \beta)\times (k^2)^{\frac{1}{2}-\alpha-\beta}
\end{equation}
where $I_{c2}(\alpha, \beta)$ reads:
\begin{equation}
    I_{c2}(\alpha, \beta)=\frac{2 \sin (\pi  \alpha ) \Gamma (1-2 \alpha ) \sin (\pi  \beta ) \Gamma (1-2 \beta ) \sin (\pi  (\alpha +\beta )) \Gamma (2 \alpha +2 \beta -1)}{\pi ^2}
\end{equation}
if $\beta<\frac{1}{2}$, $\alpha<\frac{1}{2}$ but $\alpha+\beta>\frac{1}{2}$. Otherwise, the integral diverges.

Another useful integral is the following:
\begin{equation}
    I_3(\alpha, \beta,\gamma, k, q)= \int \frac{d l}{2 \pi} \frac{1}{\left(l^2\right)^\alpha} \frac{1}{\left((l+k)^2\right)^\beta} \frac{1}{\left((l+q)^2\right)^\gamma}
\end{equation}
which has convergent result if $\beta<\frac{1}{2}$, $\alpha<\frac{1}{2}$, $\gamma<\frac{1}{2}$ but $\alpha+\beta+\gamma>\frac{1}{2}$. Otherwise, the integral diverges. 

It turns out that the closed form of this integral is not straightforward to find. However, for our purposes it will be enough to consider $I_3(\alpha, \beta,\gamma, k, k)$ which reads:
\begin{equation}
    I_3(\alpha, \beta,\gamma, k, k)=I_{c3}(\alpha, \beta,\gamma)\times (k^2)^{\frac{1}{2}-\alpha-\beta-\gamma}
\end{equation}
where $I_{c3}(\alpha, \beta,\gamma)=I_{c2}(\alpha, \beta+\gamma)$.

\section{Anomalous dimension due to Regge bounded interactions}
\label{app: anomalous dimension}
In this appendix, following the discussion in section \ref{subsec: O(N)}, we provide the expressions for the anomalous dimensions acquired by double-trace operators in the different sectors of the $O(N)$ CFT.  

First, due to the quartic interaction, the anomalous dimensions take the following closed form:
\begin{equation}
    \begin{split}
    (\gamma_{\Phi^4})^{\mbf a}_n
    &=\bigg\{
    \frac{(N+2) 4^{\Delta _{\phi }-1} \Gamma \left(n+\frac{1}{2}\right) \Gamma \left(\Delta _{\phi }+\frac{1}{2}\right)^3 \Gamma \left(n+\Delta _{\phi }\right)^2 \Gamma \left(n+2 \Delta _{\phi }-\frac{1}{2}\right)}{3 \pi  \Gamma (n+1) \Gamma \left(\Delta _{\phi }\right) \Gamma \left(2 \Delta _{\phi }-\frac{1}{2}\right) \Gamma \left(n+\Delta _{\phi }+\frac{1}{2}\right)^2 \Gamma \left(n+2 \Delta _{\phi }\right)},\\
    &\quad
    \frac{2^{2 \Delta _{\phi }-1} \Gamma \left(n+\frac{1}{2}\right) \Gamma \left(\Delta _{\phi }+\frac{1}{2}\right)^3 \Gamma \left(n+\Delta _{\phi }\right)^2 \Gamma \left(n+2 \Delta _{\phi }-\frac{1}{2}\right)}{3 \pi  \Gamma (n+1) \Gamma \left(\Delta _{\phi }\right) \Gamma \left(2 \Delta _{\phi }-\frac{1}{2}\right) \Gamma \left(n+\Delta _{\phi }+\frac{1}{2}\right)^2 \Gamma \left(n+2 \Delta _{\phi }\right)},\\
    &\quad 0
    \bigg\}.
    \end{split}
\end{equation}

The anomalous dimensions arising from the derivative interaction take the form:
\begin{equation}
\begin{split}
&(\gamma_{\partial^2 \Phi^4})^{\mbf a}_n
=\\
&\bigg\{
\frac{(N-1) 4^{\Delta _{\phi }-1} \Gamma \left(n+\frac{1}{2}\right) \Gamma \left(\Delta _{\phi }+\frac{1}{2}\right)^3 \left(\Delta _{\phi } \left(4 \Delta _{\phi }+12 n-1\right)+3 n (2 n-1)\right) \Gamma \left(n+\Delta _{\phi }\right)^2 \Gamma \left(n+2 \Delta _{\phi }-\frac{1}{2}\right)}{3 \pi  \Gamma (n+1) \Gamma \left(\Delta _{\phi }+1\right) \Gamma \left(2 \Delta _{\phi }+\frac{1}{2}\right) \Gamma \left(n+\Delta _{\phi }+\frac{1}{2}\right)^2 \Gamma \left(n+2 \Delta _{\phi }\right)},\\
&\quad
\frac{4^{\Delta _{\phi }-1} \Gamma \left(n+\frac{1}{2}\right) \Gamma \left(\Delta _{\phi }+\frac{1}{2}\right)^3 \left(\Delta _{\phi } \left(-4 \Delta _{\phi }-12 n+1\right)+3 n (1-2 n)\right) \Gamma \left(n+\Delta _{\phi }\right)^2 \Gamma \left(n+2 \Delta _{\phi }-\frac{1}{2}\right)}{3 \pi  \Gamma (n+1) \Gamma \left(\Delta _{\phi }+1\right) \Gamma \left(2 \Delta _{\phi }+\frac{1}{2}\right) \Gamma \left(n+\Delta _{\phi }+\frac{1}{2}\right)^2 \Gamma \left(n+2 \Delta _{\phi }\right)},\\
&\quad
-\frac{2^{2 \Delta _{\phi }-1} \Gamma \left(n+\frac{3}{2}\right) \Gamma \left(\Delta _{\phi }+\frac{1}{2}\right)^3 \Gamma \left(n+\Delta _{\phi }\right) \Gamma \left(n+\Delta _{\phi }+1\right) \Gamma \left(n+2 \Delta _{\phi }+\frac{1}{2}\right)}{\pi  \Gamma (n+1) \Gamma \left(\Delta _{\phi }+1\right) \Gamma \left(2 \Delta _{\phi }+\frac{1}{2}\right) \Gamma \left(n+\Delta _{\phi }+\frac{1}{2}\right) \Gamma \left(n+\Delta _{\phi }+\frac{3}{2}\right) \Gamma \left(n+2 \Delta _{\phi }\right)}
\bigg\}.
\end{split}
\end{equation}

For the exchange contributions, we have
\begin{equation}
(\gamma_{{\tt exc}})_n^{\mbf a}=-\frac{\beta_n^{\mbf a|A}(2\Delta_\phi-1)}{a_{n}^{\tt gff}}\,.
\end{equation}

For $n=0$ and $n=1$, these take the explicit forms:
\begin{equation}
\begin{split}
(\gamma_{{\tt exc}})_0^{\mbf a}
&=\left\{0,0,\frac{1-\Delta _{\phi }^2}{2 \left(2 \Delta _{\phi }+1\right)}\right\},\\
(\gamma_{{\tt exc}})_1^{\mbf a}
&=\bigg\{
\frac{(N-1) \left(4 \Delta _{\phi }+1\right) \left(\Delta _{\phi }^2-1\right)}{4 \Delta _{\phi } \left(2 \Delta _{\phi }+1\right)^2},\\
&\quad
-\frac{\left(4 \Delta _{\phi }+1\right) \left(\Delta _{\phi }^2-1\right)}{4 \Delta _{\phi } \left(2 \Delta _{\phi }+1\right)^2}, \frac{18-3 \Delta _{\phi } \left((\Delta _{\phi }+2)(4 \Delta _{\phi }-1)\Delta _{\phi }^2+\Delta _{\phi }-4\right)}{8 \Delta _{\phi } \left(2 \Delta _{\phi }+1\right)^2 \left(2 \Delta _{\phi }+3\right)}
\bigg\}.
\end{split}
\end{equation}
\section{Regge Limit of Witten diagram}
\label{App: Regge limit}
In this section, we show how to compute the $u$-channel Regge limit of the
exchange Witten diagram in $AdS_2$, following~\cite{Costa:2012cb}. We begin with
the Mellin representation of a generic Witten diagram \footnote{In one dimension, the two cross-ratios are not independent, which implies that the two Mellin variables are also not independent. Nevertheless, one can define the Mellin representation by starting from its higher-dimensional counterpart and then setting $z=\bar z$ \cite{Ferrero:2019luz}.},
\begin{equation}
    W(z)=\int [ds][dt]\,
    z^{2s} (1-z)^{2t}\,
    \rho_{\Delta_\phi}(s,t)\, M(s,t),
\end{equation}
where
\begin{equation}
    \rho_{\Delta_\phi}(s,t)
    =\Gamma^2(\Delta_\phi-s)\,
     \Gamma^2(-t)\,
     \Gamma^2(s+t).
\end{equation}
We first approximate the Gamma functions appearing in the measure for large
$s$ and $t$, retaining the first two terms using the asymptotic formula
\begin{equation}
    \Gamma(a+i x)\Gamma(b-i x)
    \sim
    2\, i^{a+b+1} e^{-\pi (x+i b)}
    \left(
        -\frac{\pi (a-b)(a+b-1)}{2 x^2}
        +\frac{\pi}{i x}
    \right)
    x^{a+b}.
\end{equation}
In addition, we make the replacement
\begin{equation}
    z^2 \;\longrightarrow\; z^2 e^{-2\pi i}.
\end{equation}
Since we are interested in extracting the $O(z^0)$ and $O(1/z)$ terms, we keep
only the leading two contributions. We then perform the $s$- and $t$-integrals
to extract the corresponding coefficients. Below, we present the Regge limits
of the spin-$0$ and spin-$1$ exchange Witten diagrams:
\begin{align}
    W^u_{\Delta,0}(z)
    &=\frac{i\pi\, \Gamma\!\left(\Delta_\phi-\tfrac{1}{2}\right)^2}
    {2\Delta z\, \Gamma(\Delta_\phi)^2
     -2\Delta^2 z\, \Gamma(\Delta_\phi)^2},
    \\ \nonumber
    W^s_{\Delta,1}(z)
    &=W^t_{\Delta,1}(z)
    =-\frac{i\pi^{3/2}\, 4^{1-2\Delta_\phi}
    \Gamma\!\left(2\Delta_\phi-1\right)^2}{z},
    \\ \nonumber
    W^u_{\Delta,1}(z)
    &=\frac{i\pi^{3/2}\, 2^{3-4\Delta_\phi}
    \Gamma\!\left(2\Delta_\phi\right)^2}
    {\Delta-\Delta^2}.
\end{align}
The $1/z$ term is absent in $W^u_{\Delta,1}(z)$.

\section{Regge superbounded basis in presence of global symmetry}
\label{App: Regge super basis}

In this section, we describe the procedure for constructing superbounded
functionals using exchange and contact Witten diagrams. We begin by defining
the tensor structures
\begin{align}
   T^S_{ijkl} &= \delta_{ij}\delta_{kl}, \nonumber\\
   T^T_{ijkl} &= \frac{1}{2}
   \left(\delta_{il}\delta_{jk}+\delta_{ik}\delta_{jl}
   -\frac{2}{N}\delta_{ij}\delta_{kl}\right), \nonumber\\
   T^A_{ijkl} &= \frac{1}{2}
   \left(\delta_{il}\delta_{jk}-\delta_{ik}\delta_{jl}\right).
\end{align}
We can then expand the four-point correlator as
\begin{equation}
\begin{split}
  \langle \phi_i(x_1)\phi_j(x_2)\phi_k(x_3)\phi_l(x_4)\rangle
  =&\;
  T^S_{ijkl}\sum_\Delta a^S_{\Delta}\,
  W_{\Delta}^{S(s)}
  +\text{t-channel }(j\leftrightarrow l)
  +\text{u-channel }(j\leftrightarrow k)
  \\
  &+
  T^T_{ijkl}\sum_\Delta a^T_{\Delta}\,
  W_{\Delta}^{T(s)}
  +(j\leftrightarrow l)
  +(j\leftrightarrow k)
  \\
  &+
  T^A_{ijkl}\sum_\Delta a^A_{\Delta}\,
  W_{\Delta}^{A(s)}
  +(j\leftrightarrow l)
  +(j\leftrightarrow k).
\end{split}
\end{equation}
Rewriting everything in the $T^{S/T/A}_{ijkl}$ basis, we obtain expressions of
the form
\begin{equation}
    \sum_\Delta a^S_{\Delta} \, PW_{\Delta}^{S|S}(z)
    + \sum_\Delta a^T_{\Delta} \, PW_{\Delta}^{S|T}(z)
    + \sum_\Delta a^A_{\Delta} \, PW_{\Delta}^{S|A}(z),
\end{equation}
with analogous expressions for the $T$ and $A$ sectors. We denote this object by $PW^{|}_{\Delta}(z)$ to emphasize that, in practice, it is not necessary
to work with the actual Polyakov blocks. Instead, one may work directly with
exchange Witten diagrams, which leads to a slightly cleaner implementation.
This is equivalent, since Polyakov blocks differ from exchange Witten diagrams
only by the addition of contact terms, whose effect is simply to shift the
coefficients with which the contact terms appear in the superbounded basis.

To make the functional superbounded, we add two contact terms together with an
exchange Witten diagram in the antisymmetric channel with
$\Delta=2\Delta_{\phi}-1$. This leads to the structure
\begin{equation}
\begin{split}
  & T^S_{ijkl}\left(\lambda_1 C^S_1+\lambda_2 C^S_2\right) \\
  &+ T^T_{ijkl}\left(\lambda_1 C^T_1+\lambda_2 C^T_2\right) \\
  &+ T^A_{ijkl}
  \Big(\lambda_2 C^A_2
  +\lambda_3\big(PW_{2\Delta_{\phi}-1}^{A(s)}
  +(j\leftrightarrow l)
  +(j\leftrightarrow k)\big)\Big).
\end{split}
\end{equation}
Here $C^b_j$, with $b=S,T,A$, denote the components of the two allowed contact
terms, and we note that $C^A_1=0$. Writing everything in the
$T^b_{ijkl}$ basis, we then impose superboundedness in each of the three sectors.
This results in a total of nine constraints, corresponding to the cancellation
of both the $O(z^0)$ and $O(1/z)$ terms.

Although this may appear overconstraining, only two crossing equations are
independent, and as a result we find a unique solution for the unfixed
coefficients multiplying the contact terms. To illustrate this, consider the
singlet sector. One finds that $PW^{i|S}_{\Delta}$ behaves as $1/z$ for
$i=S,T,A$. The $O(z^0)$ term is canceled by contributions from the two-derivative
contact term and the antisymmetric exchange diagram, thereby fixing
$\lambda^S_3$. Canceling the $1/z$ terms in $PW^{i|S}_{\Delta}$ then yields three
constraints for the two remaining undetermined coefficients
$\lambda^S_1$ and $\lambda^S_2$. As noted above, only two of these constraints are
independent, which uniquely determines both coefficients. This procedure fixes $\lambda^i_1$, $\lambda^i_2$, and $\lambda^i_3$ in all
sectors. In particular, $\lambda_3$ corresponds to the
$\tilde{\omega}_3$ functional, which possesses the desired superboundedness
property. We have provided the three independent functionals that we find explicitly in a
\textsc{Mathematica} file submitted with this version of the paper. We note that,
by decomposing the Witten diagrams into conformal blocks with double-trace
exchanges and their derivatives, one in fact obtains a full basis of
superbounded functionals. However, only three of these are independent in
addition to the set of Regge-bounded functionals.

\bibliography{bib}
\bibliographystyle{JHEP}

\end{document}